\documentclass[a4paper,fleqn]{cas-dc}

\usepackage[authoryear,longnamesfirst]{natbib}
\usepackage{pdflscape}
\usepackage{lscape}
\usepackage{rotating}
\usepackage{amssymb}
\usepackage{float}

\usepackage{graphicx}
\usepackage{subcaption}

\usepackage{booktabs}
\usepackage{threeparttable}
\def\tsc#1{\csdef{#1}{\textsc{\lowercase{#1}}\xspace}}
\tsc{WGM}
\tsc{QE}
\tsc{EP}
\tsc{PMS}
\tsc{BEC}
\tsc{DE}

\begin{document}
\let\WriteBookmarks\relax
\def\floatpagepagefraction{1}
\def\textpagefraction{.001}

\shorttitle{TSTEM: A Cognitive Platform for Collecting Cyber Threat Intelligence in the Wild}

\shortauthors{P. Balasubramanian et~al.}

\title [mode = title]{TSTEM: A Cognitive Platform for Collecting Cyber Threat Intelligence in the Wild}                      


%
\author[1]{Prasasthy Balasubramanian}[type=editor,
                        auid=000,bioid=1]
                        \ead{prasasthy.balasubramanian@oulu.fi}






\affiliation[1]{organization={Center for Ubiquitous Computing, University of Oulu},
    addressline={Pentti Kaiteran katu 1 }, 
    city={Oulu},
    postcode={90570}, 
    country={Finland}}

\author[1]{Sadaf Nazari}
\ead{sadaf.nazari@oulu.fi}
\author[1]{Danial Khosh Kholgh}

\ead{danial.khoshKholgh@oulu.fi}


\credit{Data curation, Writing - Original draft preparation}


\author [1]{Alireza Mahmoodi}
\ead{abakhshi21@student.oulu.fi}


\author%
[1]
{Justin Seby}
\ead{justin.seby@oulu.fi}


\author%
[1]
{Panos Kostakos}
\cormark[1]
\ead{panos.kostakos@oulu.fi}


\cortext[cor1]{Corresponding author}



\begin{abstract}
The extraction of cyber threat intelligence (CTI) from open sources is a rapidly expanding defensive strategy that enhances the resilience of both Information Technology (IT) and Operational Technology (OT) environments against large-scale cyber-attacks. However, for most organizations, the collection of actionable cyber threat intelligence remains both a technical bottleneck and a black box. While previous research has focused on improving individual components of the extraction process, the community lacks open-source platforms for deploying streaming CTI data pipelines in the wild. To address this gap, the study describes the implementation of an efficient and well-performing platform capable of processing compute-intensive data pipelines based on the cloud computing paradigm for real-time detection, collecting, and sharing CTI from different online sources. We developed a prototype platform (TSTEM), a containerized microservice architecture that uses Tweepy, Scrapy, Terraform, ELK, Kafka, and MLOps to autonomously search, extract, and index IOCs in the wild. Moreover, the provisioning, monitoring, and management of the TSTEM platform are achieved through infrastructure as a code (IaC). Custom focus crawlers collect web content, which is then processed by a first-level classifier to identify potential indicators of compromise (IOCs). If deemed relevant, the content advances to a second level of extraction for further examination. Throughout this process, state-of-the-art NLP models are utilized for classification and entity extraction, enhancing the overall IOC extraction methodology. Our experimental results indicate that these models exhibit high accuracy (exceeding 98\%) in the classification and extraction tasks, achieving this performance within a time frame of less than a minute. The effectiveness of our system can be attributed to a finely-tuned IOC extraction method that operates at multiple stages, ensuring precise identification of relevant information with low false positives.

\end{abstract}



\maketitle

\section{Introduction}
The convergence of Information Technology (IT) and Operational Technology (OT) has led to significant changes in various industries and supply chains, including energy, manufacturing, food, medicine, and transportation. While this convergence has brought many benefits, it has also increased the potential attack surface for cyber threats \cite{9784874,Fortinet2021,filkins2019sans}. Recent incidents, such as Stuxnet, Colonial Pipeline, and Kaseya VSA, have demonstrated the far-reaching impacts that cyber attacks can have on critical infrastructure and various communities and industries. To address this growing security challenge, the real-time collection and sharing of actionable cyber threat intelligence (CTI) have become a critical defense approach for enhancing the security posture of connected places, spaces, businesses, and organizations \cite{zrahia2018threat,kokkonen2022autonomy,anagnostopoulos2021challenges}.

Cyber threat intelligence, which covers a  broad and fragmented ecosystem of vendors and tools for collecting, analyzing, and disseminating information on possible cyber threats and vulnerabilities, is estimated to be worth 1,6 billion \cite{va2022frost}. The CTI market is divided into two main segments: Threat Intelligence Platforms (TIPs) and Cyber Threat Intelligence (CTI). TIPs are software tools that help organizations collect, analyze, and disseminate threat intelligence data, while CTI refers to the actual intelligence gathered and analyzed to understand the nature and origins of cyber threats. TIPs are often integrated with other security tools (e.g., firewalls, CTI, intrusion detection systems, and SecOps) in order to provide a comprehensive view of an organization's security posture. In short, the growth of threat sharing between organisations reflects both the widening threat landscape introduced by the IT/OT convergence and the proliferation of connected devices generating an unprecedented amount of heterogenous data \cite{mehmood2019implementing}.

Artificial Intelligence and Machine Learning approaches are also playing an important role in producing faster and better threat telemetry as data sets grow and become more complex and heterogeneous. Notably, AI is used predominately in the following CTI tasks: i) automated threat analysis and triage, ii) malware analysis and classification, iii) anomaly detection and intrusion prevention, iv) network traffic analysis, v) predictive modeling and forecasting of cyber threats, vi) natural language processing (NLP) and text mining of unstructured data sources. Furthermore, recently we have seen the heightened emphasis on automation tools (e.g., automated report generation and summarization) that can enable workflows and trigger playbooks to mitigate threats automatically \cite{mavroeidis2022cybersecurity}.
\paragraph{Motivation:} Cyber threat intelligence (CTI) entails the systematic acquisition of information aimed at proactively safeguarding an organization's digital infrastructure and assets. This information is assimilated by end-user systems, such as firewalls, Intrusion Detection Systems (IDS), and Threat Intelligence Platforms (TIPs), which obtain the data from third-party providers through threat feeds. These feeds are accessible via diverse storage solutions, including data warehouses, data lakes, or RESTful Application Programming Interfaces (APIs), and are available through either public or paid services. The atomic Indicator of Compromise (IOC) represents the most prevalent and extensively employed form of cyber threat intelligence (CTI) in the field. Atomic IOCs contain immutable artifacts like IP addresses, hashes, URLs, domain names, etc., which are used to detect and mitigate possible risks and to plan strategies and tactics to avoid or respond to attacks \cite{hutchins2011intelligence}. 

Presently, the techniques utilized for detecting Indicators of Compromise (IOCs) present a significant obstacle for many organizations, primarily due to the insufficiency of tools required to efficiently gather these artifacts from vast data repositories, websites, and social media platforms found "in the wild". \cite{vlachos2022saint,borges2022methodological}. Consequently, defence systems often depend on third-party vendors and community feeds (e.g., \cite{team,spamhaus,shadowserver}) to ingest these indicators, which renders the data-collecting process a black box with potentially severe security flaws due to the absence of provenance and transparency in the data collection \cite{mitra2021combating,ranade2021generating}. There are various challenges to overcome.
\paragraph{Challenge 1:} General-purpose web crawlers that harvest all links and index the HTML pages are suboptimal solutions for domain-specific searches and threat-hunting tasks. Respectively, parsers that only extract context from a static seed file are limited by their inability to explore a wider crawling frontier dynamically. Prior research has attempted to fill this gap by using vanilla focused crawlers that target specific domains. However, these crawlers have been limited by the fact that they do not incorporate the latest state-of-the-art (SOTA) large language models (LLMS). Therefore, a significant required advancement in the CTI field is the development of focused crawlers with high-performing AI capabilities trained with data from the cybersecurity domain.  In summary, there is a requirement for a platform that can effectively crawl designated websites, gather the content in a structured manner, and incorporate an AI cognitive module to examine, display, and extract relevant IOCs from the acquired data.
\paragraph{Challenge 2:}  The underutilization of advanced machine learning (ML) and deep learning (DL) models in CTI solutions is a critical challenge as it limits the potential of these tools to enhance their capabilities. Effectively, existing research and frameworks in the CTI literature tend to rely heavily on basic machine learning and deep learning algorithms, as well as rule-based algorithms, without explicitly incorporating cutting-edge algorithms for extracting actionable CTI. This approach may limit the accuracy and effectiveness of CTI solutions, as it prevents more sophisticated analysis of data, which could lead to increased precision, relevance and discovery of new zero-day threat indicators in the wild.
\paragraph{Challenge 3:} Most platforms are provided to consumers as a service (AAS) or as API endpoints, and they are typically designed to either gather data and analyze it offline or to manually extract information from data obtained offline. So, another challenge is the need for open-source platforms that enable near real-time end-to-end analytics pipelines that improve transparency on how and what data are collected from open-source data. 
\paragraph{Contribution:} The study presents a novel approach for IOCs discovery, extraction and indexing using transformer-based algorithms and demonstrates the deployment of threat intelligence infrastructure as a code (IaC). Therefore, the main contributions of the paper are as follows:
\begin{enumerate}
    \item Novel AI-based focus crawlers using custom domain language models for clear, deep, and social web. 
    \item A novel containerized architecture to handle computationally demanding data pipelines based on the cloud computing paradigm.
    \item Real-time detection, collecting, and sharing of CTI from different online sources. 
    \item An innovative approach for transparent and automated focus crawling, indicator of compromise (IOC) extraction, and indexing using Transformer-based algorithms.
    \item Demonstration of using infrastructure as code for deploying CTI tools.
\end{enumerate}

\section{Related Work}
In recent years, the digital forensics field as a whole and the CTI domain specifically have experienced considerable advancements, leading to the creation of numerous tools designed to facilitate and enhance the investigative process. These tools can be categorized into four primary groups: Desktop Forensic, which includes tools like ProDiscover Basic, Encase, Recuva, Cyber Check Suit, and Autopsy; Live Forensic, featuring tools such as Volatility Framework, Win-Lift, Belkasoft, Magnet RAM, and OSF Mount; Network Forensic, with tools such as Wireshark, Ettercap, Nmap, Nessus, Snort, and Suricata; and finally, OSINT forensic, which comprises theHarvester, Maltego, and Spiderfoot \cite{kanta2020survey, lovanshi2019comparative}. This section examines the most significant related work in the realm of cyber threat intelligence, specifically focusing on open-source intelligence acquisition.

\subsection{AI powered Crawlers}
Web scraping and crawling are essential to the development of cyber threat intelligence, which is a prominent approach for preventing cyber attacks. Despite significant advances in data processing techniques with the help of AI, the most recent research on web crawling for CTI has yet to reach the state-of-the-art in AI utilization. Most web crawling studies use rule-based or basic algorithms like Term Frequency–Inverse Document Frequency (TF-IDF) for ranking the pages or filtering data. Research studies for extracting IOCs from Twitter feeds also appeared not to achieve state-of-the-art using AI.

Niakanlahiji et al. \cite{niakanlahiji2019iocminer} developed a framework called IoCMiner for extracting IOCs from the Twitter feed. IoCMiner consists of two subsystems: the CTI Expert Finder (CTIEFinder) and CTI Extraction subsystems. CTIEFinder finds the CTI experts who publish cyber threat information, and the tweets posted by these experts are then passed through CTI Tweet Classifier. They use a Random Forest classifier to filter relevant tweets, which uses bags of words tokenization and then passes through the IOC extractor module, which is a rule-based logic. They were able to achieve 97\% accuracy with the complete process. 

In a separate study involving Twitter, Behzadan et al. \cite{behzadan2018corpus} used a convolutional neural network (CNN) based binary classifier to identify relevant tweets and then fed the results into a multi-class classifier to determine the type of cyber threat information contained in the tweet. According to the study, the binary classifier had an accuracy of 94.72\%, while the threat type detection classifier had an accuracy of 87.56\%. Dion{\'\i}sio et al. \cite{dionisio2020towards} used a multi-task learning framework for cyber threat detection from Twitter data. The researchers established a pipeline for experimenting with CNN/RNN/Bi-LSTM-based embedding for a binary classifier to identify relevant tweets. Furthermore, a Conditional Random Field (CRF) layer was utilized to generate Named Entity Recognition (NER) stage predictions, which facilitated the extraction of IOCs. They achieved a maximum of 97.9\% accuracy for classification and 97.3\% for entity extraction models. 
In \cite{tundis2022feature}, proposed a method to automate the assessment of threat intelligence sources and predict their relevance using a model based on meta-data and word embedding. Regression models are trained to predict the relevance score of sources on Twitter. Results claimed that the assigned score could reduce waiting time for intelligence verification and improve early threat detection.

Multiple studies focus on web crawling and scrapping that uses components with AI algorithms. Bamboat et al. \cite{bamboat2022web} provides an overview of web mining techniques and explores web content mining techniques such as wrapper generation, page content mining, and web crawlers, including their classification and the tools used. The researchers examined fundamental techniques to using the bag-of-words method for vectorizing, followed by the application of classification methods. Similarly, Sharma et al. \cite{sharma2022web}  recently published a survey paper on web page ranking based on various web mining techniques, which explains the different algorithms used for content analysis from web pages, including sequential, classification, and clustering methods. The algorithms listed for classification techniques comprise Decision Tree, Bayesian, basic Artificial Neural Networks, SVM, etc.

Guojun et al. \cite{guojun2017design} proposed an advanced  intelligent dynamic crawler that stores XPath data extraction rules in a database, dynamically loads the rules depending on the target, and uses TF-IDF methods to compute relevance. In a separate study with a focus on cybersecurity, Jenkins et al. \cite{jenkins2021designing} implemented a framework that consolidates data into an easily searchable software platform (i.e., Splunk). This platform enables the indexing of large volumes of cybersecurity-specific data. The data collection process utilizes a web crawler and web scraper, which rely on a set of 200 keywords to gather relevant information. Subsequently, the indexed data becomes readily accessible and useful for research, analysis, and development within the cybersecurity domain.

Nathezhtha et al. \cite{nathezhtha2019wc}, implemented Web Crawler based Phishing Attack Detector (WC-PAD), which uses web traffic, web content, and Uniform Resource Locator (URL) as input and leverages a rule-based algorithm to classify to detect a phishing attack. WC-PAD uses a Domain Name System (DNS) blacklist to identify the credibility of the web address in the first phase and heuristic analysis for detecting phishing attacks.

Pham et al. \cite{pham2018phishing} implemented a neuro-fuzzy framework (dubbed Fi-NFN), a combination of the artificial neural network and the fuzzy logic to classify phishing URLs to do anti-phishing on fog networks. They have used feature extraction and some basic neural network architectures as part of their implementation. In a more recent study, Kanneganti et al. \cite{kannegantirecurrent} used an RNN (Recurrent Neural Network) to create a real-time text classification system with web crawling for identifying open data sets on the internet.

As previously noted, there is a lack of focused crawlers with high-end AI capabilities. The majority of crawlers examined thus far predominantly exhibit generic characteristics, leading to the extraction of substantial amounts of irrelevant data from general web pages. The subsequent section delves into existing research that discusses the development and architectures of focused crawlers, addressing this critical gap in the field. 

\subsection{Focused Crawlers}
Nunes et al. \cite{nunes2016darknet} developed a focused crawler to collect data from Darknet/Deepnet markets and forums. They implemented a binary classifier to detect the relevance of the web page's content and trained it to recognize products and topics related to drugs, weapons, etc., that are not relevant to malicious hacking from the web page content. In the study published in \cite{charles2020focused}, the authors developed a search engine with a focused crawler that logically uses lexical semantics and architecture with Breadth First Search (BFS) and Depth First Search (DFS) techniques. This study primarily focuses on the design of the focused crawler architecture and does not include steps for analyzing and exploring the data.

In \cite{bozkir2023grambeddings} a novel deep neural model for phishing URL identification is proposed. The proposed model employs n-gram embeddings calculated in real-time, eliminating the need for pre-training or word/sub-word level information, thus reducing computational demands. The model includes an adjustable and automated n-gram selection and filtering mechanism and a neural network architecture with cascading CNN, LSTM, and attention layers. They achieved an accuracy of of 98.27\% in their study.

Web scraping and crawling are crucial to the development of cyber threat intelligence, which is a vital approach to preventing cyber attacks. While data processing techniques have advanced significantly due to artificial intelligence (AI), most current web crawling studies still rely on rule-based or basic algorithms like Term Frequency–Inverse Document Frequency (TF-IDF) for ranking pages or filtering data. Similarly, research studies on extracting indicators of compromise (IOCs) from Twitter feeds have yet to utilize AI fully. However, some studies \cite{dionisio2020towards} have shown promising results using AI-based approaches, including convolutional neural networks, and multi-task learning frameworks achieving accuracies of up to 97.9\% for classification and 97.3\% for entity extraction. Multiple studies have also demonstrated the potential for combining web crawling and scraping with AI algorithms, including decision trees, Bayesian algorithms, basic artificial neural networks, and support vector machines. Despite these advancements, there remains ample opportunity for enhancing the application of AI in web crawling and scraping with the ultimate goal of advancing cyber threat intelligence development.

\subsection{CTI platforms with AI capability}
There has been a significant amount of research on integrating artificial intelligence and machine learning with cyber threat intelligence to obtain actionable threat telemetry from a range of sources. However, many of these studies have employed basic techniques, indicating a need for further advancement in this field.

Dutta et al. \cite{dutta2020overview} used a Naïve Bayes classifier to identify threats in data collected from various sources, including both structured (STIX, CyBOX, TAXII) and unstructured (hacker's forums, blogs, blacklists) sources. The study faced limitations due to the utilization  s of basic  machine learning algorithm and the absence of in-depth data analysis for extracting genuine indicators of compromise. Ghazi et al. \cite{ghazi2018supervised} used a named entity recognition technique in natural language processing (NLP) and regex parsing to identify IOCs from various unstructured sources. They implemented the Conditional Random Fields (CRFs) algorithm as a named entity recognition (NER) model. 

Several studies have explored the use of knowledge graphs for cyber threat intelligence. Mittal et al. \cite{mittal2019cyber}, introduced Cyber-All-Intel, a pipeline for extracting knowledge in the form of vector spaces, representing this knowledge using knowledge graphs, and performing analytics on the data in the cybersecurity informatics domain. In addition to its VKG (Vector space Knowledge Graph) structure, which combines knowledge graphs and vector embeddings elements, this pipeline is also equipped with a query engine and an alerting system that analysts can use to uncover actionable insights. This provides analysts with a comprehensive tool for knowledge extraction, representation, and analysis in the cybersecurity informatics.

The authors in \cite{al2022cyber} combine DNN Deep Neural Network (DNN)—a multi-layered artificial neural network capable of autonomously capturing latent data patterns—with PCA (Principal Component Analysis), a technique used for reducing data dimensionality while retaining key features. Additionally, they employ statistical and knowledge-based methods to enhance threat detection in an intrusion detection system (IDS). The method merges the DNN model and PCA to achieve the target, claiming to use low computing power and resources. 

Vlachos et al. \cite{vlachos2022saint} presented a framework called SAINT (Systemic Analyzer In Network Threats), which extracts data from various information sources such as Twitter, the clear net, bug bounty programs, and the deep web. The authors used regular expression analysis, entity analysis, time series descriptive analysis, and statistical analysis to extract cyber threat intelligence (CTI) from these sources. This framework aims to provide a comprehensive approach to CTI analysis by collecting and processing data from a variety of sources and using various techniques to extract and analyze relevant information.

Koloveas et al. \cite{koloveas2021intime} developed a platform called "inTime", which includes a crawler, CTI extractor, and sharing platform. The framework is based on the ACHE crawler, an open-source framework with built-in basic ML algorithms for topic classification. InTime also has a social media monitoring component that uses Twitter data collected using the Twitter API and a basic classifier with a Random Forest and CNN architecture. The CTI extraction module of inTime consists of a pre-trained natural language processing (NLP) module called spaCy, which is used as a Python library with several built-in ready-to-use tools. 

A study by Zhao et al. \cite{zhao2020timiner} introduced TIMiner, an automated framework designed for CTI extraction and sharing, leveraging social media data. TIMiner employs a convolutional neural network to identify the targeted domain of CTIs and utilizes an IOC extraction approach grounded in word embedding and syntactic dependence for the identification of new IOC types. The experimental results of this study demonstrated promising accuracy levels for both the CTI domain recognizer and IOC extraction (84\% and 94\%, respectively).

Guarascio et al. \cite{guarascio2022boosting} developed a platform called ORISHA (ORchestrated Information SHaring and Awareness) with an interface to a honeynet Intrusion Detection System (IDS). They employed bagging and boosting ensembles in combination with decision tree models as weak learner classifiers, performing multiple iterations. Additionally, they utilized active learning methods to incorporate unlabeled data for iterative training of machine learning models. After five iterations, the average F-Measure increased from the initial value of 0.870 to 0.971. 

The paper of Skarmeta et al. \cite{10.1007/978-3-031-36096-1_4} presents a platform based on the Malware Information Sharing Platform (MISP) for the automatic exchange and processing of threat information. They are leveraging Federated learning scheme as part of their platform with multi-layer Artificial Neural Network and have achieved around 80\% accuracy. The paper explains the implementation of a permissioned blockchain platform for auditing, incorporating access control through distributed Identity Management (IdM), and employing Privacy-Enhancing Technologies (PETs) like K-anonymity, L-diversity, and T-closeness, along with differential privacy to safeguard sensitive data during sharing.

Menges et al. \cite{menges2021dealer} propose a decentralized platform called DEALER for the exchange of threat intelligence information. This platform not only facilitates compliance with legal reporting obligations for security incidents but also provides additional incentives for information exchange among involved parties. The authors implement the platform based on the EOS blockchain and IPFS distributed hash table, demonstrating its feasibility and cost-efficiency through a prototype and cost measurements. The DEALER platform operates on a cryptocurrency-based incentive system, ensuring pseudonymous exchange of structured incident information. Verifiers use objective CTI quality indicators to establish incident reputation, aiding buyers in selecting relevant incidents. Dispute resolution mechanisms and cryptocurrency incentives protect buyers and sellers. While presenting a fully decentralized model for sharing Cyber Threat Intelligence (CTI), the authors emphasize the platform's design considerations for legal and privacy requirements. They highlight the need for future work to ensure privacy and compliance, particularly with regulations like the General Data Protection Regulation (GDPR).

Preuveneers et al. \cite{preuveneers2020distributed} enhance their existing security framework TATIS, with distributed ledger capabilities for reliable and auditable threat intelligence sharing, implementing and evaluating it on MISP with real-world threat feeds. The study also highlights the importance of trust and responsible information sharing in their work. 
Overall, the research on this topic suggests that there is potential for further improvement by incorporating state-of-the-art artificial intelligence-natural language processing (AI-NLP) algorithms, such as transformer-based models. These advanced techniques could enhance the field's effectiveness and efficiency. For conciseness, Table \ref{tab:Related work CTI platforms with AI1} summarizes some of the studies discussed in this review, along with the datasets, algorithms/methods, and evaluation scores/efficiency that they achieved.

{\tiny
\begin{table*}

\caption{Summary of core papers for CTI extraction with AI -Part-1}
\label{tab:Related work CTI platforms with AI1}       
\centering
\begin{tabular}{p{0.7in} p{1.4in} p{1.5in} p{1.1in} p{1.3in}}
\hline\noalign{\smallskip}
Citation & Data source & Method & Evaluation & Features \\
\noalign{\smallskip}\hline\noalign{\smallskip}
\cite{dutta2020overview} & They have used a combination of latest CVE entries of 2020 and CWE list v4.1 along with malware dataset and Goodware dataset from ‘kaggle’ web portal. &  Naïve Bayes algorithm for the classification task.  & The accuracy of the model reflected as 98.2\% and 96.6\% for the training dataset and test dataset respectively. & Threat detection classifier model\\
\cite{ghazi2018supervised} & CTI documents from FireEye, Kaskpersky Security Lab, and a curated list of APT reports from a Github repository. & Natural language processing to extract threat feeds from unstructured cyber threat information sources. & A 70\% precision rate has been achieved. & Entity recognition model for extracting IOCs\\
\cite{mittal2019cyber} & Corpus is collected from  chat rooms, dark web, blogs, RSS feeds, social media, and vulnerability databases. The current corpus has 85,190 CVEs from the NVD dataset, 351,004 Tweets, and 25,146 Reddit and blog posts.
& Vectorized Knowledge Graph (VKG) representation and query engine to get the results  from the data.  & Achieved  81.5\% accuracy. & Query engine based on Knowledge graph\\
\cite{al2022cyber} &  PCAP files from CIC IDS 2018 dataset, SE CIC IDS 2018 and the ISCX IDS 2012 and CIC IDS 2017 dataset. & PCA is applied primarily in dimensionality reduction. binary classification. & Model achieves 98\% accuracy rate while SOTA achieved 97\%. & Binary classifier model for threat detection\\
\cite{kannegantirecurrent}&Trained using 300 web pages which were collected after evaluation of the amount of text, size, and genre of the web page.  &Text classifier for web page contents. &Precision has been optimized to a level of  85\%. \\
\cite{tanvirul2022cyner}& 60 threat intelligence reports referenced in the MITRE attack website under the software category, where each threat report describes a unique malware. & Entity recognition model. & Precision 75.30\%, recall 78.07\% and F1 Score 76.66\%. & Entity recognition model for extracting IOCs \\
\cite{vlachos2022saint} & The study uses various methods to collect and analyze data related to cyber threats and attacks. Social Network Analysis SNA and Clear Net Crawler CNC are used to collect data from Twitter and open internet sources respectively. BBA Bug Bounties Analyzer is used to analyze bug bounty programs and rewards, while DWC deep web crawler is used to explore black markets and unregulated forums. & Regular Expression Analysis, Entity Analysis, Time Series Descriptive Analysis, Statistical Analysis. & The study does not mention any specific evaluation scores. Combining SNA and CNC can leverage their strengths to detect both known and emerging cyber-attacks. & Extract CTI based on different methods\\ 

\noalign{\smallskip}\hline
\end{tabular}
\end{table*}
}

\begin{table*}

\centering
\begin{tabular}{p{0.7in} p{1.4in} p{1.5in} p{1.1in} p{1.3in}}
\hline\noalign{\smallskip}
Citation & Data source & Method & Evaluation & Features \\
\noalign{\smallskip}\hline\noalign{\smallskip}
\cite{nunes2016darknet} & The data is collected from two sources on the darknet/deepnet: markets and forums.  & Naive Bayes (NB), random forest (RF), support vector machine (SVM) and logistic regression (LOG-REG). & Attained a recall of 92\% and a precision rate of 82\%. & Binary classifier model to detect the relevance of web-page contents\\ 
\cite{koloveas2021intime} & Tweets collected with IoT vulnerabilities and CVE tags (Common Vulnerabilities and Exposures), along with MITRE CVE dictionary. & Word2vec embedding with the content ranking technique for web crawlers and pre-trained model spaCy for CTI extraction. Also, use Twitter data classifier for social media component developed with CNN and Random forest algorithm followed by the CTI extractor. & 95\% accuracy achieved for Twitter content classifier.  & Platform for CTI extraction and sharing with a classifier model along with spacy entity extraction\\
\cite{zhao2020timiner} &The data collected from security blogs (AlienVault, FireEye, Webroot, etc),
security vendor bulletin (Microsoft, Cisco, Kaspersky, etc) and the posts published in hacking forums (Webroot, HackerForum, etc). & The proposed method uses convolutional neural network (CNN) based recognizer to identify the domain of a CTI, and a hierarchical IOC extraction method. The CTI domain recognizer uses a word2vec model to embed threat descriptions, and applies convolutional operations to learn the features of CTIs in different domains.  &  The CTI domain recognizer and IOC extraction have demonstrated accuracy levels of 84\% and 94\% respectively. & CTI Domain Classifier model and rule-based IOC extractor.\\
 \cite{guarascio2022boosting} & CICIDS2017 dataset, a well-known IDS benchmark provided by Canadian Institute of Cybersecurity, containing several up-to-date cyberattacks & The proposed method uses bagging and boosting ensembles in combination with decision tree models as weak learner classifiers for identifying threat from PCAP files obtained from honeynet IDS &  Demonstrated improvement in F-score from 0.870 to 0.971 with multiple iterations & Orchestration information sharing platform with integrated ML models and active learning components.\\
 \cite{10.1007/978-3-031-36096-1_4} & Developed a synthetic financial dataset for fraud detection using PaySim, a tool that simulates mobile money transations based on an original dataset &  Federated learning scheme with multi-layer Artificial Neural Network &  Achieved around
80\% of accuracy & Federated learning scheme along with PETs to safeguard sensitive data. \\ 
 TSTEM (our platform) & We collected web pages and tweets, annotating them for positive and negative cases. We used the dataset from \cite{tanvirul2022cyner} for entity recognition. & Our method uses multiple-step extraction of IOCs and separate models for sentence-level and page-level classification. We are using SOTA LLMs BERT, Longformer and BERT-NER & We have evaluated our sentence classification, page classification and NER models based on Precision, Recall and F1-score. Also verified the final output IOCs on various platforms like VirusTotal and Alienvault & A platform featuring autonomous infrastructure deployment, which integrates collection, extraction, and indexing capabilities by harnessing the benefits of MLOps, the Infrastructure as Code (IaC) methodology, and state-of-the-art models like BERT, Longformer, and BERT-NER. \\
\noalign{\smallskip}\hline
\end{tabular}
\end{table*}
\subsection{Autonomous Deployment of Infrastructure}
A significant proportion of studies in the field of data crawling and cyber threat intelligence are constructed using a RESTful web service architecture. Some studies like \cite{vlachos2022saint}, discussed in the above section, have implemented their platform based on REST API service. Similarly, \cite{koloveas2021intime} have also developed their platform components based on REST API service. 

Borges et al.\cite{borges2022methodological} created a tool tailored to meet the demands of cybersecurity analysts, enabling them to inject, analyze, and schedule threat data for the purpose of gaining insights and enhancing their ability to contextualize threats effectively. Their results show that this helps to understand the context in which threats are placed, making vulnerability mitigation more effective. The tool was developed using Python and Flask, web development micro-frameworks that enable efficient and robust applications. The analysis steps in this study were implemented using the JavaScript language and the D3.js library. This allows the tool to work with large datasets in the feed and create complex data-driven graphics and dynamic visualizations of IOCs.

Opara et al. \cite{opara2022auto} conducted experiments using AutoML on three different cloud platforms (Microsoft Azure, Google, and IBM) to generate rules and models for detecting threats with cybersecurity data. Automated Machine Learning (AutoML), is the process of using built-in models that are part of the cloud platform to automate the process of training and deploying an ML model to the cloud platform. In the study mentioned, the multi-label classifier achieved up to 70\% accuracy, with the IBM platform performing the best. In addition to accuracy, the study also incorporated platform-based evaluation metrics, such as Mean Absolute Percentage Error (MAPE), Root Mean Square Error (RMSE), and Mean Absolute Error (MAE), as crucial components of their assessment. Within this context, the cloud service provider is responsible for the maintenance and development of the underlying infrastructure.

Leitner et al. \cite{leitner2020ait} also used the Infrastructure as a Code (IaC) methodology to set up a flexible Cyber range for exercises, training, and research activities. Their study incorporated the MLOps (Machine Learning Operations) methodology,  aiming to optimize the interaction between data and machine learning while standardizing the transition of models from development to production environments. MLOps is an automation pipeline or framework for deploying machine learning models in production efficiently and reliably. 

Despite existing advancements, a research gap persists in creating a robust framework that combines MLOps for machine learning and deep learning models with the Infrastructure as Code methodology for efficient deployment and maintenance of the entire framework. Addressing this gap offers a valuable opportunity for further exploration in this area. Overall, the research efforts examined in this section highlight the growing interest in the research community to automate the collection of CTI from diverse OSINT sources and apply this information to enable prompt alerts on emerging threats. Nevertheless, only a small number of studies have investigated the automation of CTI infrastructure deployment. Moreover, there is limited research focusing on the deployment of pipelines for OSINT collection using fine-tuned generative models, which could facilitate full autonomy within the system.

\section{Methodology}
This study proposes a framework for web crawling and data collection from a wide array of sources, encompassing social media, the dark web, and the surface web. The collected data is then preprocessed and analyzed using a combination of state-of-the-art artificial intelligence, natural language processing algorithms and rule-based algorithms, depending on the type of data (structured or unstructured). The ultimate goal of this process is to deploy an autonomous platform for extracting indicators of compromise in the wild and at scale. This section presents an overview of the different data sources and the functions of the modules developed in this implementation.
\subsection{CTI sources explored}
\paragraph{Twitter:} We use Twitter as the source of CTI extraction from social media data. Twitter is a treasure trove of free data generated by users worldwide via tweets that contain text, videos, photos, or links. 
Twitter hosts various professional communities where users share their recent news and opinions through their Twitter accounts. One particular community of interest within the realm of cybersecurity is the group of individuals who dedicate their time and effort to voluntarily sharing the latest indicators of compromise with the rest of the world. These individuals are actively engaged in the cybersecurity community and play a valuable role in helping to protect against threats by disseminating relevant information and insights. Prior research \cite{dionisio2020towards} indicates that a substantial portion of this data is not listed on the popular blacklists, which was one of our motivations for using the Twitter platform as the social media for data in this OSINT application.

The primary component of our API is a Crawling module that utilizes the Twitter API streaming feature to continuously extract tweets containing indicators of compromise (IoCs) from live stream data. This module is accompanied by a CTI extractor, which includes an AI-driven classifier and an IoC extractor module responsible for extracting IoCs from relevant tweets. This API allows for the real-time collection and analysis of data from social media sources (i.e., Twitter), enabling the real-time identification of potential threats.

\paragraph{Dark/Clear Web:} The surface web (commonly known as the clear web) is another valuable source for gathering relevant data in the field of cybersecurity. Many security experts and cybersecurity groups actively share their findings through blog posts and data troves (e.g., IP dumps) on various platforms and websites, providing a wealth of information for those interested in staying up-to-date on the latest threats and best practices. These can be utilized for extracting valuable threat intelligence with the shortest delay.

 Apart from the traditional web pages that exist primarily on the surface of the internet and are searchable via search engines (e.g., Google and Bing), there is also the "dark" web. The dark web is essentially a part of the internet that is not indexed by search engines and, therefore, is hidden from most users. Specifically, the dark web is a term used to describe the content on the World Wide Web that is only accessible via the darknet, which is an overlay network utilizing the internet. The darknet comprises both small peer-to-peer networks and more expansive popular networks, including Tor, Freenet, I2P, and Riffle, which are managed by public organizations and individuals. Tor, also known as Onion, is one such network that utilizes a traffic anonymization technique by routing onions under the network's top-level domain suffix ".onion."

Searching the dark web for relevant information is arguably more challenging than the clear web. As a result, we have decided to incorporate a Dark Web API and a Clear Web API as a source of information for collecting IoCs. Our Dark/Clear Web API utilizes a web crawling algorithm to access and analyze \textit{.onion} sites on the dark web, or regular URLs on the clear web. The module then applies a page classifier only to retrieve content from pages that contain CTI information before extracting only the relevant information with an IOC extractor.
\subsection{Data collection pipelines of TSTEM}
The quality of threat data is crucial for businesses to prevent cyber attacks. However, the process of collecting this data through scraping and crawling the internet comes with numerous challenges. The components of the end-to-end pipelines deployed and tested by the TSTEM are as follows. 

\paragraph{Kafka:}The live stream data is collected and forwarded to Kafka, a distributed streaming platform for publishing and subscribing to streams of records. Kafka is an event-streaming platform that is frequently used in high-performance data pipelines, streaming analytics, data integration, and mission-critical applications. It is particularly well-suited for use in microservices due to its scalability, flexibility, efficiency, and high speed. It also facilitates communication between services while maintaining incredibly low latency and fault tolerance. We have used two main components of Kafka in our application: Kafka producer and consumer.

\paragraph{Kafka Producers and Consumers:} In the Twitter application, the module handling the Twitter API acts as a producer, a client application that publishes events to Kafka. Producers are responsible for writing events to Kafka. The same logic applies to the clear and dark web modules. The consumers subscribe to (read and process) these events from Kafka; the cognitive modules, powered by multiple SOTA AI-NLP algorithms, work as consumers here. In our application, we have leveraged the fully decoupled and agnostic nature of producers and consumers in Kafka to achieve high scalability. This allows our application to handle a large volume of data without sacrificing performance. For example, producers in Kafka do not need to wait for consumers, which improves the throughput time of the producer and helps the consumer to work offline though the final effect would be real-time processing without compromising on the performance of each module. We have used the benefits of Kafka to make the data crawling tasks more robust and efficient. The data is then indexed to Elasticsearch for further reference and processing. 

\paragraph{Elastic Stack:}The Elastic Stack, comprised of Elasticsearch, Logstash, and Kibana (ELK), is a suite of open-source tools designed for centralized logging and problem identification in servers and applications. By enabling searches of all logs in a single location and facilitating connections between logs from multiple servers over a specified time period, the Elastic Stack facilitates efficient issue resolution. Elasticsearch serves as a repository for logs, while Logstash handles log shipping and processing. Kibana, a web interface featuring visualization capabilities, may be hosted on either Nginx or Apache.

\paragraph{Twitter Pipeline:}Twitter is a widely-used social media platform with a daily active user base of 100 million and generates 500 million daily tweets. These tweets can potentially contain various types of IOCs. In this module, we utilize the Twitter API to crawl the live stream of tweets. The Twitter API provides extensive access to public Twitter data, and allows users to manage and provide access to their own non-public Twitter information, such as Direct Messages, to developers for further analysis. This API enables read and write capabilities for Twitter data, allowing for the composition of tweets, access to profile information, and the retrieval of high volumes of tweets on specific topics or in specific locations.

\begin{figure}[h]
\centering
  \includegraphics[width=0.50\textwidth]{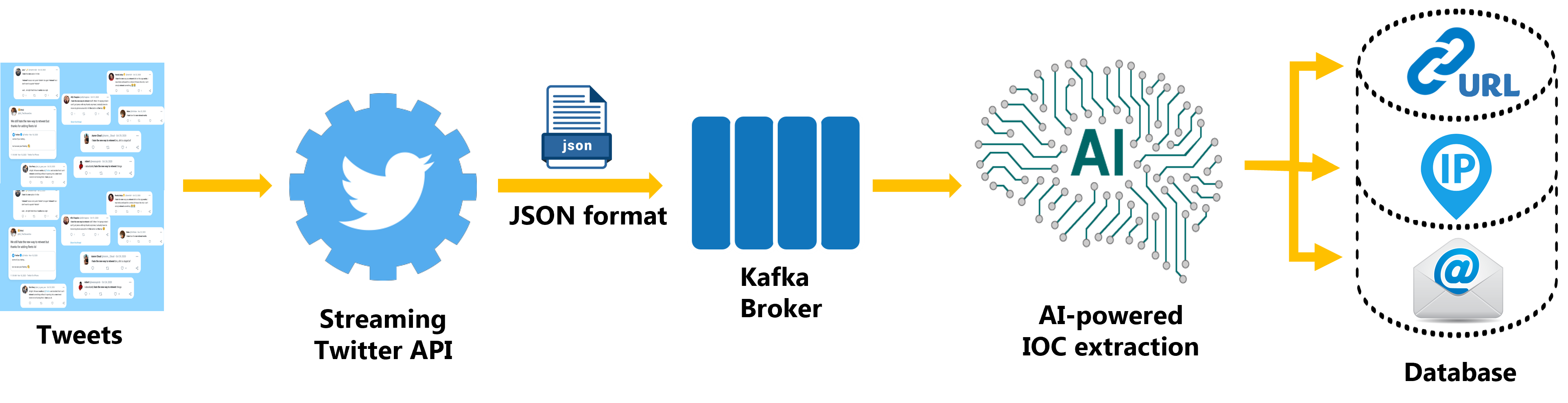}
\caption{Dataflow used to train the AI model to detect IOC from streaming tweets in real-time.}
\label{fig:Twitter_fig1}       
\end{figure}

\begin{figure}[h]
\centering
  \includegraphics[width=0.50\textwidth]{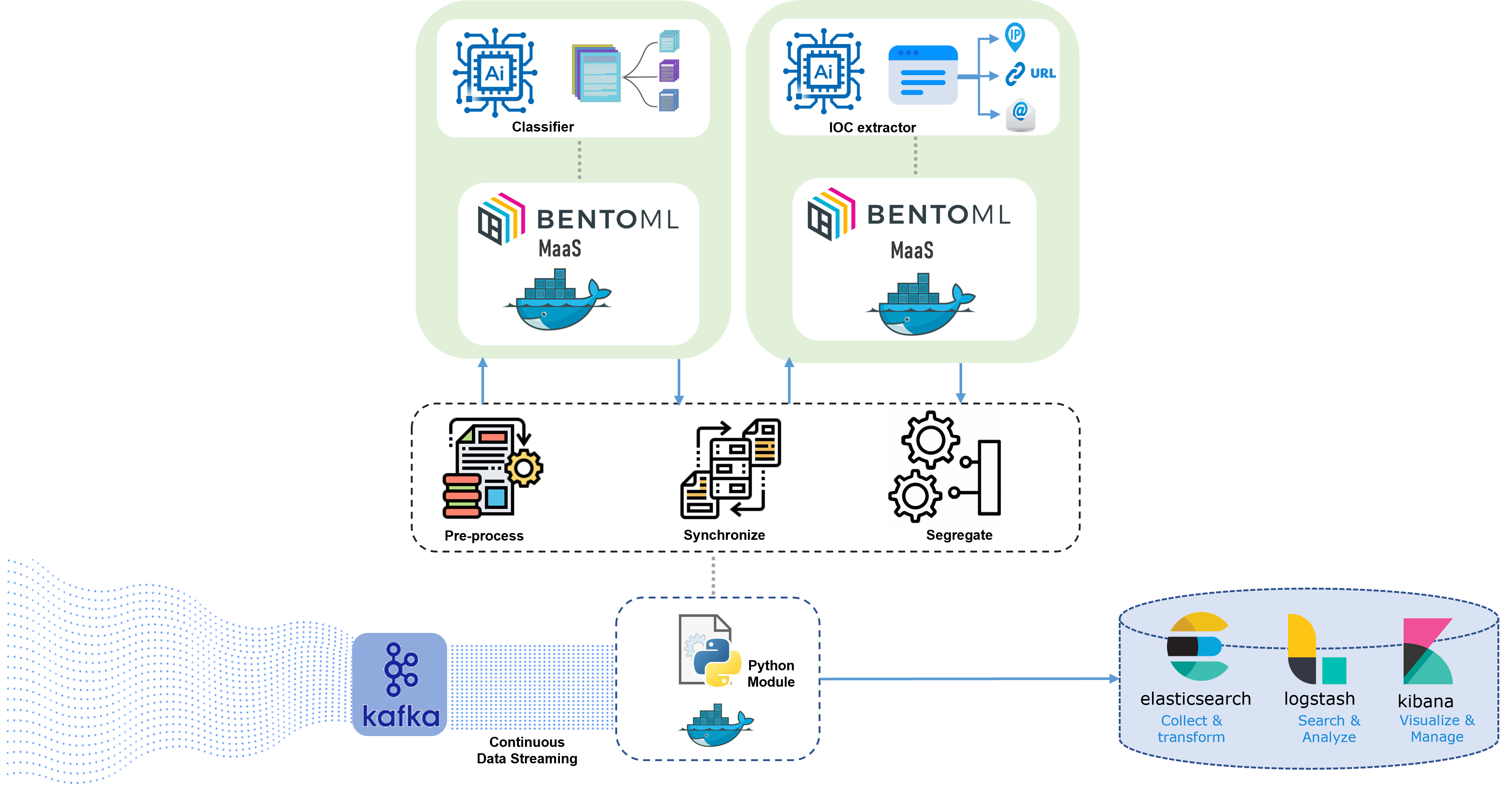}
\caption{Dataflow of shared components and their interactions.}
\label{fig:Shared_components}       
\end{figure}

The cognitive module of the Tweet Crawler consists of multiple components powered by AI-NLP algorithms, one for detecting the relevance of tweet data and the other for extracting IoCs from the relevant tweets. Once the relevance is identified, a rule-based and entity-recognition AI algorithm is used to extract IOCs from the relevant tweets. The final list of IOCs is generated from the combination of output from both rule-based and an entity recognition AI algorithm and is then indexed to Elasticsearch using the full Elastic Stack cluster. The dataflow of streaming tweet processing in real-time is as shown in Figure\ref{fig:Twitter_fig1}. Also, Figure \ref{fig:Shared_components} represents the general dataflow between the shared components and the web crawler.
\begin{figure}[h]
\centering
  \includegraphics[width=0.50\textwidth]{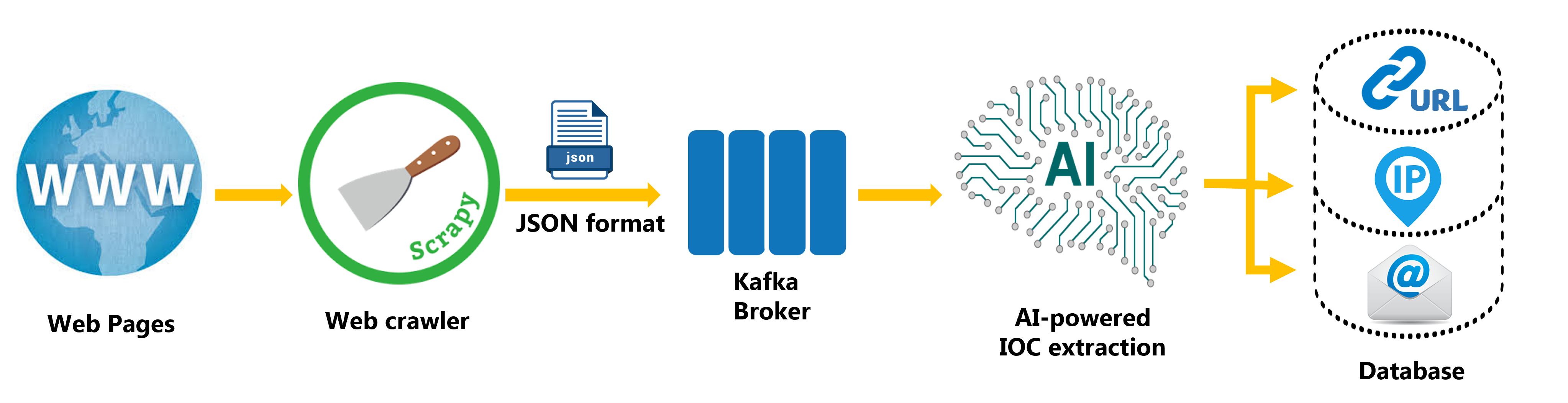}
\caption{Dataflow used to train the AI model to detect IOC from streaming web pages in real-time.}
\label{fig:Web_fig2}       
\end{figure}
\paragraph{Dark/Clear Web Pipelines:} Previously mentioned, active communities and security research groups consistently disseminate intelligence to the broader community through cybersecurity websites and blog posts. Additionally, darknet sites may contain various relevant IOCs and other leaked, encrypted information. As expected, the initial stage of this component involves collecting data from web pages on the internet through web crawling.
A web crawler, also known as a spider or bot, is a tool that allows users to specify custom behavior for the purpose of crawling and analyzing the pages of a specific website (or group of websites). We are using multiple spiders classified based on the seed files used, and they are ACHE, Sitemap, Ahmia, Wiki, etc. These spiders are user-defined as custom classes per the use cases' requirements. The custom class includes parameters such as the URLs to be crawled, the data to be extracted, and crawling frequency. The spiders are then launched using Scrapy, an open-source python framework to extract data using APIs. We defined crawlers for the dark web and the Clear web in a separate Scrapy framework and launched them as independent microservices for crawling.
\paragraph{Focus Crawling:}As part of the study, we have implemented focused crawlers with brute-force logic incorporated with a lexical semantic algorithm using regular expressions in python. The method is implemented to avoid repeated visits to the same web pages and unnecessary data processing.

After extracting data from websites, AI-powered page classification is used to determine whether a site is suitable for extracting indicators of compromise. If a page is classified as relevant, a rule-based and entity-recognition (NER) AI algorithm is employed to collect all available metrics on the page. Dark web crawlers uses specific spiders to crawl darknet sites, while clear web crawler uses different spiders to crawl clearnet sites. Both frameworks include a crawler, a page classification module, and an IOC extraction module (consisting of a combination of a rule-based algorithm and entity recognition). The dataflow of streaming web data processing in real-time is shown in Figure \ref{fig:Web_fig2}. Figure \ref{fig:Shared_components} represents the dataflow among the shared components with the Twitter framework. The subsequent section will discuss the utilization of microservices in the underlying high-level architecture of both the dark web and the clear web.

\subsection{Architecture and Implementation}
 \paragraph{CTI Module:}In our research, the CTI module operates in two distinct phases. Initially, it evaluates the relevance of the gathered data for potential IOC extraction. Once this relevance is ascertained, the data undergoes further processing in the extraction module, utilizing a hybrid approach that combines an NER model with a rule-based algorithm to extract IOCs.

 \paragraph{Language Models:} In our CTI module, we adopted two specific algorithms to classify threat data based on its type and volume. We have used BERT (Bidirectional Encoder Representations from Transformers) \cite{devlin2018bert} for short text classification, which is one of the state-of-the-art language models in natural language processing (NLP) and is also suitable for Twitter data. We have used Longformer \cite{beltagy2020longformer} for web page classification, which is appropriate for large text classification. BERT uses the Transformer architecture, which incorporates an attention process that learns the contextual associations among words within a text. A Transformer consists of two distinct components: an encoder that interprets the textual input, and a decoder that generates a task-specific prediction.
 
BERT is an encoder-focused model because it is a language model in which each masked prediction depends on other tokens in the set received at the encoder level, removing the decoder layer's necessity. BERT is a deep bidirectional model, which means it simultaneously learns information from left to right and right to left.
 
 The BERT framework has a two-step process that involves pre-training and fine-tuning stages. Pre-training involves training the model on a large dataset, while fine-tuning involves adapting the model to a specific task or dataset. This allows BERT to achieve state-of-the-art results on a wide range of natural language processing tasks. We have used a pre-trained model, which can be further fine-tuned for performing specific functions  such as text classification. Table \ref{tab:Data_sources_used} represents details of the data we have used in our training stage for each model. 
\begin{table}
\caption{Data sources used}
\label{tab:Data_sources_used}
\centering
\begin{tabular}{ llll p{0.4cm} } 
\hline\noalign{\smallskip}
Source & Task & Train & Test & Val. \\
\noalign{\smallskip}\hline\noalign{\smallskip}
Twitter & Sentence Classification & 25000& 1000 & 600 \\
Web & Page Classification & 1116 & 138 & 124 \\
Twitter/Web & NER & 4085 & 294 & 785 \\
\noalign{\smallskip}\hline
\end{tabular}
\end{table}
\paragraph{Tweet Classification:} The text classification task is used to determine the relevance of a streaming tweet from the Twitter API. Fine-tuning is a process that involves adding an untrained feed-forward layer on top of the pre-trained BERT model, using domain-specific data from the field of Cyber Security. This allows the model to adapt to the specific task or dataset and achieve superior performance. There are several benefits to fine-tuning a pre-trained BERT model. One of the main advantages is the ability to leverage transfer learning. Since a pre-trained BERT model has already been trained on a large dataset, it has already learned a significant amount of semantic and syntactic information about the language.
Therefore, it takes less time to train a fine-tuned model. Another is that less data is required. Using a pre-trained BERT requires minimal task-specific fine-tuning, so less information is specific to the cybersecurity domain without compromising performance. To fine-tune a pre-trained BERT model to identify relevant tweets in the cybersecurity domain, we collected a corpus of 26,600 tweets using the method described in  \cite{niakanlahiji2019iocminer}. In our experimental setup, we adopted the random sub-sampling technique. This method is frequently employed to partition a dataset into training, testing, and validation subsets. The split of data is shown in Table \ref{tab:Data_sources_used}. The training set is utilized for training the model on extensive text data, the test set is used for evaluating its language comprehension abilities, and the validation set helps to fine-tune and optimize the model's performance. The fine-tuned BERT model was then used as an inference model to determine the relevance of streaming data in the final data pipeline.

\paragraph{Entity Recognition:} After extracting relevant tweets, they are processed through a combination algorithm comprising an entity recognition algorithm and a rule-based algorithm. The entity recognition algorithm utilized in this case is based on BERT. The BERT model is trained with IOB (Inside-Outside-Beginning) annotated data for identifying IOCs from tokenized sentence-level data. The training data was obtained from the GitHub repository of CyNER \cite{tanvirul2022cyner}, where it was utilized to train transformer-based models. We have reused the same data for fine-tuning our BERT-base-uncased NER model. A total of 5164 records were used, and the random sub-sampling method was employed to create train, validation, and test sets. The data is annotated for five classes: malware, indicators of compromise, system, organization, and vulnerability. The final list of IOCs is generated from the combination of output from both rule-based and an entity recognition AI algorithm.
\paragraph{Page Classification:} After extracting data from websites, an AI-powered page classification module determines if it is suitable for IOC extraction. The page classifier has been trained on  1378 web pages, as shown in Table \ref{tab:Data_sources_used} that have been manually tagged, with a 45-55\% split between positive and negative records.  In this instance as well, we utilized the random sub-sampling technique to divide the data into training, testing, and validation subsets. If a page is deemed relevant, the IOC extraction tool is used to collect all existing IOCs on the page. As illustrated in Figure \ref{fig:Web_fig2}, the pipeline closely resembles Twitter's, with the primary distinction being the use of web crawlers as opposed to Twitter API listeners to supply data to the CTI extractor. As mentioned before, the crawling is carried out using Scrapy, a popular web scraping library of Python. Scrapy is a fast, high-level web crawling and scraping framework that can be applied to a variety of tasks, including data mining, monitoring, and automated testing.
\paragraph{Longformer: }In the Twitter module, we utilized a transformer-based BERT language model to classify the content of tweets. However, BERT is not capable of efficiently processing long texts due to the quadratic increase in memory and time consumption. The most natural way to address this problem is by slicing the long text by a sliding window, as BERT suffers from insufficient long-range attention, or by simplifying its transformer architecture, which would need customized CUDA kernels. In simpler terms, the self-attention layer of BERT has a complexity of O(n²), where n is the length of the input sequence. This means that it becomes more computationally expensive to process longer sequences.

BERT is typically limited to processing sequences of up to 512 tokens at a time to balance the trade-off between accuracy and computational cost. Due to this limitation, we are not using the BERT classifier for the dark web and Clear web, as the amount of textual data that needs to be classified is enormous compared to Twitter data. We are using transformer-based Longformer model as a page classifier over here. 

The Longformer \cite{beltagy2020longformer} essentially combines several attention patterns like sliding window attention mechanism, dilated sliding window attention mechanism and global/Full self-attention mechanism to overcome the drawbacks of other transformer-based models for long sequence classification. By using the combination of above attention mechanisms, the Longformer model was able to outperform other transformer-based SOTA models on certain tasks like document classification and Q\&A. Longformer is pre-trained on a document corpus and fine-tuned for six tasks including classification, Q\&A, and cross-reference resolution. The resulting model is capable of processing sequences that are up to 4,096 tokens long, which is eight times longer than what can be handled by BERT. 

The Longformer is pre-trained with Masked Language Modeling (MLM). The purpose is to recover randomly masked tokens in sequence. Because MLM pre-training from scratch is costly, the model is pre-trained from RoBERTa-approved checkpoints while making only the minimal changes necessary to support Longformer's attention mechanism. This attention pattern is further plugged into a pre-trained transformer model without changing the model architecture. The Lonformer base model is trained using the following parameters used to optimize the training process and help the model learn effectively from the data \cite{beltagy2020longformer}: i) 65K gradient updates, ii) Sequence length: 4,096, iii) Batch size: 64 (218 tokens), iv) Maximum learning rate: 3e-5, v) Linear warmup of 500 steps, and vi) Power 3 polynomial decay. The rest of the hyperparameters are the same as RoBERTa. For fine-tuning, we have used a learning rate of \(2e-5\), set the number of training epochs to 5, used a weight decay of 0.01, and chose a batch size of 2, considering the high computational demands on the GPU during model training.

\begin{figure}[h]
  \includegraphics[width=0.48\textwidth,trim={0 0 0 1cm},clip]{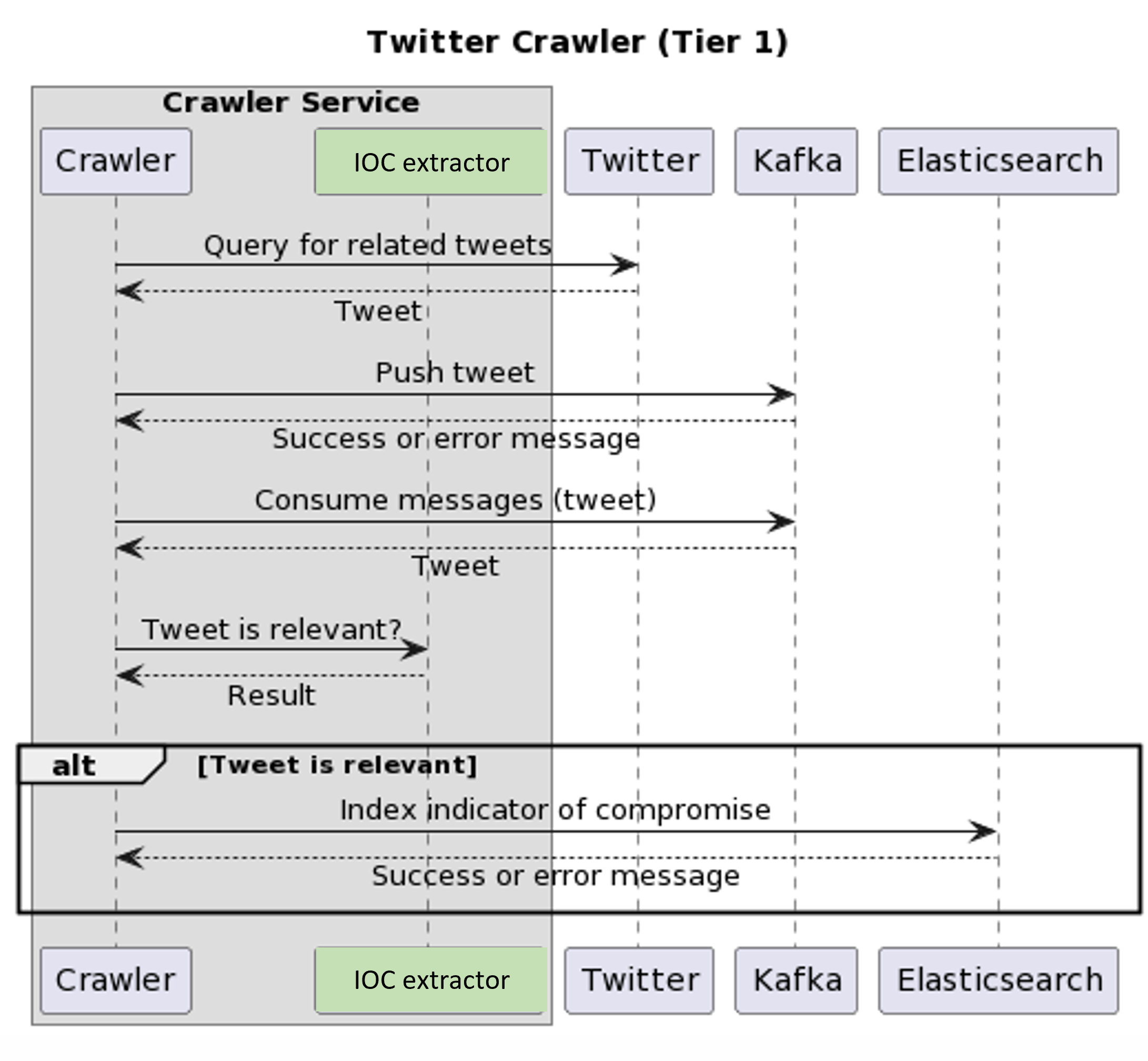}
\caption{Sequence diagram for Twitter crawler}
\label{fig:Twitter_seq}       
\end{figure}
\begin{figure}[h]
  \includegraphics[width=0.48\textwidth,trim={0 0 0 1cm},clip]{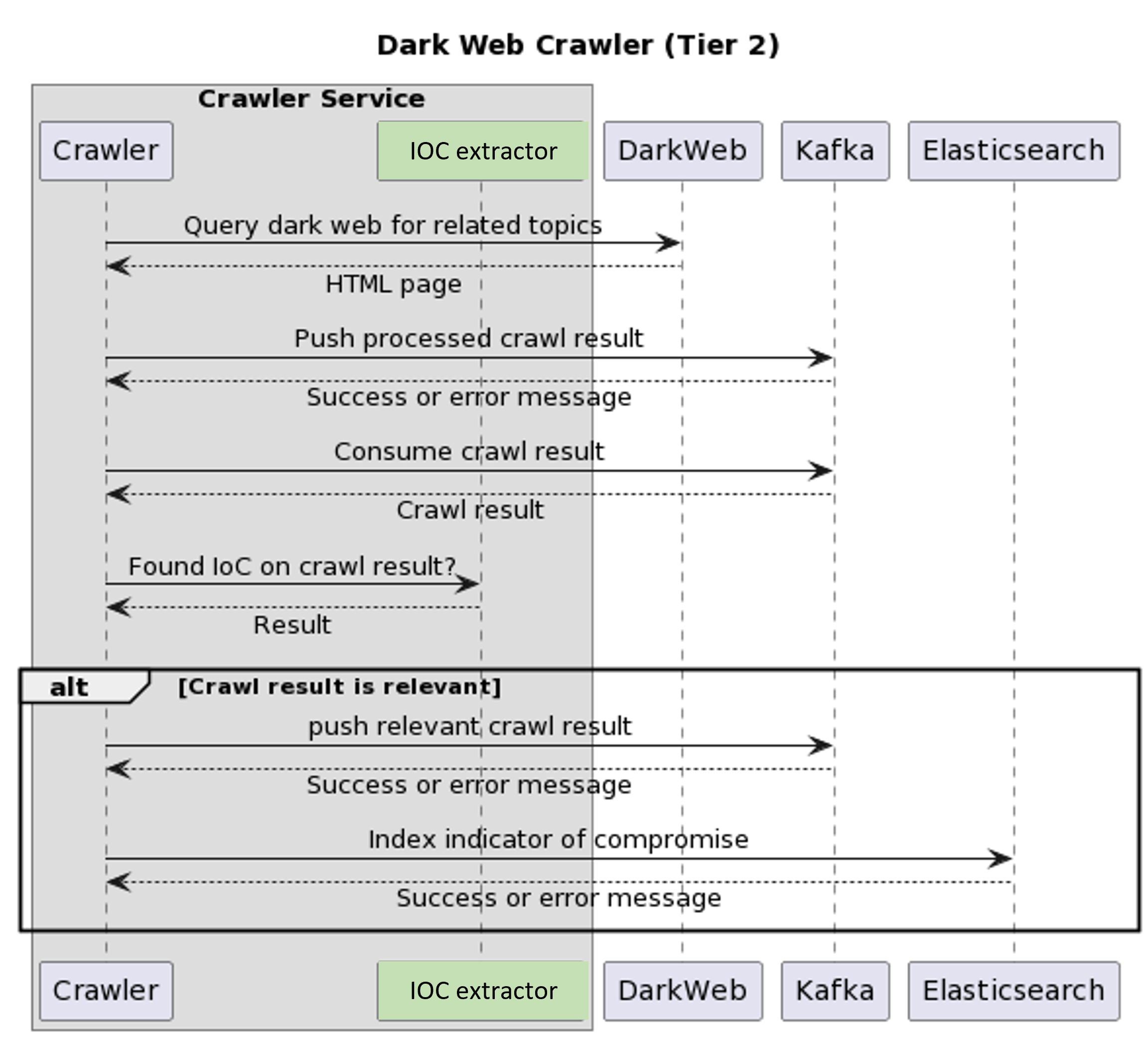}
\caption{Sequence diagram for Web crawler}
\label{fig:Web_seq}       
\end{figure}
\paragraph{Process Interactions:}Figure \ref{fig:Twitter_seq}  represents the sequence diagram for the Twitter crawler framework. The framework continuously collects data via the Twitter API. The collected data is then subjected to a classifier and NER extractor to gather IOCs from relevant information. Figure \ref{fig:Web_seq} represents a similar sequence diagram of web crawlers. The representation shows data processing sequences in the dark web and the clear web crawler framework. The data is continuously collected and processed through a classifier and extractor for collecting IOCs.
\paragraph{Integration:}The data extracted during crawling is continuously pushed to a Kafka cluster, serving as a messaging queue similar to that of the Twitter application. The CTI module of the crawlers reads data from the Kafka cluster to feed it into the page classifier. The data collection daemons and the other support components used in the data collection framework have been virtualized using docker. As shown in the sequence diagram (Fig. \ref{fig:Web_seq})
the data crawlers push data to a messaging queue (i.e., Kafka) and/or index data into Elastic Stack for downstream analytics, CTI sharing, etc. The code for the entire infrastructure is publicly available in the TSTEM GitHub repository \footnote{https://github.com/PrasasthyKB/TSTEM}.
\begin{figure*}[h]
\centering
  \includegraphics[width=1.00\textwidth]{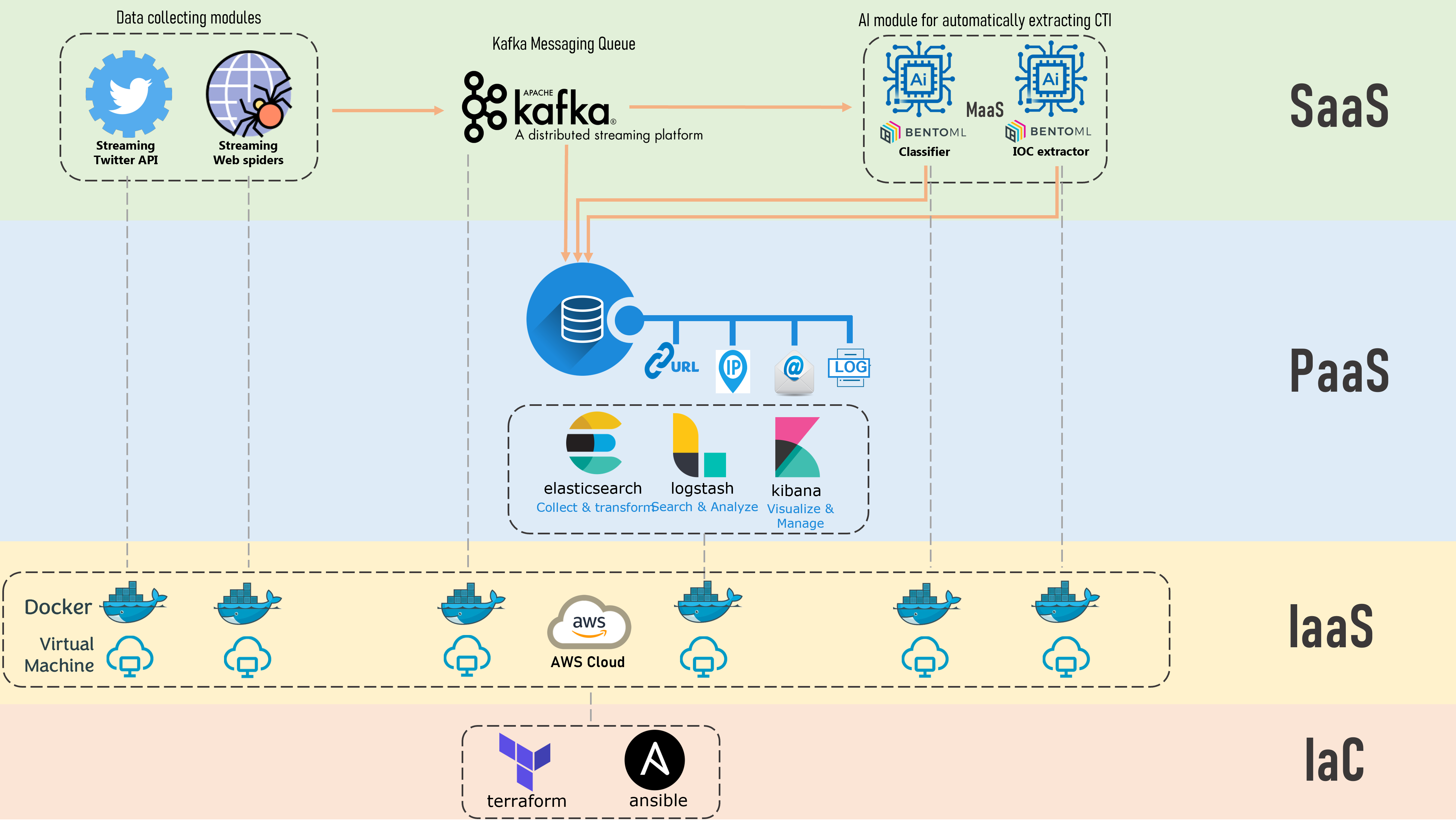}
\caption{The proposed architecture employs a microservices design for web crawling and analytics, with the option to extend the architecture to include other services through the use of an additional virtual private cloud (VPC).}
\label{fig:Microservices}       
\end{figure*}
\paragraph{Infrastructure:}The infrastructure is provisioned as code using Terraform, enabling the deployment of multiple daemons and services on the same Virtual Private Cloud (VPC) interface and across different VPCs and networks seamlessly. This approach makes the entire framework highly low-maintenance. Furthermore, encapsulating the data collection daemons in docker containers provides robust isolation of the runtime environments. The Infrastructure as a Code (IaC) deployment enables seamless access to easy update mechanisms, faster migrations, and improved server consolidation.Additionally, Terraform enables scalability by using a declarative configuration language to define infrastructure, allowing for modular organization and parameterization of resources. Through variables, modules, and dynamic scaling techniques, Terraform facilitates the creation and management of scalable infrastructure. Its state management capabilities and integration with cloud provider services further streamline the process, enabling efficient scaling based on evolving requirements. As part of MLOps, we deploy all deep learning models as Models as a Service (MaaS) by provisioning with BentoML. BentoML is an open-source framework that makes deploying ML-powered prediction services easy, ready to deploy, and scalable in a CI/CD/CT (Continuous Integration/Continuous Deployment/Continuous Testing) pipeline. It helps accelerate and standardize deploying ML/DL models in a robust format as part of the final product. BentoML facilitates ML model scalability by packaging models as containerized microservices, allowing for independent scaling of specific functionalities. Its integration with container orchestration platforms, such as Kubernetes, enables dynamic scaling based on workload demands. Additionally, BentoML supports versioning, automated deployment through CI/CD pipelines, and REST APIs, providing a streamlined approach for the scalable and efficient deployment of machine learning models in production.
\paragraph{Cloud Computing Resources:} The microservice architecture of the crawling mechanism is presented in Figure \ref{fig:Microservices}. The specification of the VMs varies according to the workload. The Elastic Stack VM hosting Elasticsearch, Logstash, and Kibana uses compute-optimized instances with 4 vCPU and 8GiB (c6i.xlarge). A Kafka VM has 2 vCPU and 4GiB (t3.medium). Our Web crawler VM uses an optimized memory instance with 4 vCPU and 32GiB of memory (r61.xlarge), and the Web classifier VM also uses a compute-optimized instance with 8 vCPU and 16GiB of memory (c5.2xlarge). Finally, the Twitter crawler VM has 8 vCPU and 32GiB (t3.2xlarge). The specification written here could be updated in the future following computation needs.

 \paragraph{Bill of Materials:} Below, we provide a list of the core frameworks and libraries that we have used:
\begin{enumerate}
\item Scrapy Engine: Scrapy is a fast, easy-to-use framework for web crawling and data scraping that can be used for a range of purposes such as data mining, monitoring, and automated testing.
\item Tweepy (4.6.0), which is a python library and enables us to use Twitter API.
\item Kafka-python (2.0.2) along with Flask (2.0.3), which handles large data processing and resolves bottlenecks.
\item Pytorch (1.11.0), which helps us to classify tweets (This is not mandatory to use as we have already trained our model).
\item Bentoml (0.11.0), which enables us to modularize our trained model for further use in our pipeline.
\item Iocextractor (1.13.1), which helps us to not only extract the IOC’s but also convert them to their original format using regular expression rules.
\item Elasticsearch (8.4.3), which enables us to index our final data.
\item Docker (5.0.3) to pack everything together and offer a micro-service that works independently.
\item Terraform (1.1.3) and Ansible (ansible 5.8.0, ansible-core 2.12.6) for provisioning the under-laying infrastructure.
\end{enumerate}

\section{Results}
This section presents the results of the study obtained through the research methods outlined in previous sections. Concurrently, the next section provides a discussion of the implications of the findings and any attendant limitations of the study. Furthermore, the results are contextualized within the broader landscape of previous research in the field, thus facilitating a more comprehensive understanding of the contributions of the present work.

\subsection{Models}
\paragraph{Twitter BERT Classifier:} The results of our study demonstrate the efficacy of our model for sentence-level classification on Twitter feeds. By extracting the feed at the sentence level and applying the classification model, we were able to achieve an accuracy rate of 98\%. Table \ref{tab:Evaluation_scores} represents details of evaluation scores achieved. Furthermore, the implemented model displayed notable effectiveness in distinguishing between relevant and non-relevant data streams obtained from Twitter API, with average precision and recall rates of approximately 98\%. These findings suggest that our approach holds promise for the analysis of social media data.
\begin{table}[h!]
\caption{Evaluation scores for classification of tweets.}
\label{tab:Evaluation_scores}       
\begin{tabular}{lllll}
\hline\noalign{\smallskip}
Label & Accuracy & Precision & Recall &F1-score \\
\noalign{\smallskip}\hline\noalign{\smallskip}
Relevant & 0.98 & 0.98 & 0.96& 0.97 \\
Non-Relevant & 0.98 & 0.97 & 0.99 & 0.98 \\
\noalign{\smallskip}\hline
\end{tabular}
\end{table}

\paragraph{Web Page Longformer Classifier:}
The model classifies web pages based on their content by considering data extracted at the page content level. This allows the model to make accurate predictions about the content of a web page based on its text and other features. The Model is classifying with very high accuracy of 95\%. Table \ref{tab:Longformer_Eval_scores} represents evaluation details. The model has an average precision and recall of approximately 95\% for identifying relevant and non-relevant pages.

\begin{table}[H]
\caption{Evaluation scores for web-page classification.}
\label{tab:Longformer_Eval_scores}       
\begin{tabular}{lllll}
\hline\noalign{\smallskip}
Label & Accuracy & Precision & Recall &F1-score  \\
\noalign{\smallskip}\hline\noalign{\smallskip}
Relevant & 0.95 & 0.96& 0.92& 0.94\\
Non-Relevant & 0.95 & 0.94 & 0.97 & 0.96\\
\noalign{\smallskip}\hline
\end{tabular}
\end{table}

\begin{table*}[H]
\caption{System Metrics for models}
\label{tab:System_eval}       
\centering
\begin{tabular}{p{3.5in} p{3in}}
\hline\noalign{\smallskip}
Artifacts collected & Quantitative/Qualitative measurements\\
\noalign{\smallskip}\hline\noalign{\smallskip}
 Average classification time for webpages & 31.10s\\
 Average IOC extraction time & 8.23s \\
 Ratio of relevancy in percentage & true - 10.34 / false - 89.66\\
 Average classification time for Twitter data& 3.29s\\
 Average IOC extraction time for Twitter data& 3.38s\\
 Ratio of relevancy in percentage & true - 28.1 / false - 71.9\\
\noalign{\smallskip}\hline
\end{tabular}
\end{table*}

\paragraph{NER Extraction Algorithm:}
The employed Named Entity Recognition model performs keyword extraction at the sentence level for the given dataset. This methodology encompasses segmenting the data into individual sentences and subsequently processing them through the model. As a result, the model accurately discerns and extracts relevant keywords or entities, achieving a notable accuracy rate of 98.74\%. The results in Table \ref{tab:NER_Eval_scores} show the evaluation scores for different labels in an extraction task. The precision and recall metrics indicate the model's proficiency in accurately classifying each respective label. The F1-score represents the harmonic mean of precision and recall, offering a consolidated metric that encapsulates both dimensions of performance.

Overall, the model appears to perform well, with precision and recall scores around 0.8\% for the Indicator, Malware, Organization, and System labels. The Vulnerability label has a lower F1-score of 0.48\%, indicating that the model may be less accurate at classifying this label. It is important to consider that the scores associated with individual labels may be affected by factors such as the label's frequency within the dataset and the complexity of the classification task for each label.

\begin{table}[h!]
\caption{Performance Metrics for NER Category Classification.}
\label{tab:NER_Eval_scores}       
\begin{tabular}{llll}
\hline\noalign{\smallskip}
Label & Precision & Recall &F1-score  \\
\noalign{\smallskip}\hline\noalign{\smallskip}
Indicator& 0.83& 0.85& 0.84 \\
Malware & 0.85 & 0.94 & 0.90 \\
Organization & 0.81 & 0.87 & 0.84 \\
System & 0.71 & 0.76 & 0.73\\
 Vulnerability & 0.45 & 0.5 & 0.48\\
\noalign{\smallskip}\hline
\end{tabular}
\begin{tablenotes}
\small
\item Model validation accuracy 0.98; validation loss 0.06.
\end{tablenotes}
\end{table}

\subsubsection{System Metrics for Models}
The results in Table \ref{tab:System_eval} demonstrate the performance of the TSTEM system in collecting and analyzing various types of data using our AI-NLP models. The table presents a range of quantitative and qualitative measurements, including the average classification and IOCs (indicators of compromise) extraction times for webpages and Twitter data and the relevancy ratio for each data type.

Our findings indicate that the system demonstrates comparatively rapid mean classification durations for both webpages and Twitter data. The accelerated classification time for Twitter data can be attributed to the implementation of a sentence-level classifier. Additionally, it is observed that entity extraction models exhibit analogous patterns, with extended processing times for web pages in comparison to Twitter data. In terms of relevancy, the system demonstrated a higher ratio of relevant information extracted from Twitter data compared to web pages. These findings suggest that the system effectively and autonomously identifies and extracts relevant information from different data sources.

\subsubsection{System Metrics for Crawlers}
Table \ref{tab:crawl_eval}  shows a variety of metrics linked to the performance of the system for continuously crawling webpages and tweets over a period of three days. The table displays a range of indicators, including the total number of pages and tweets crawled, the number of relevant pages and tweets found, and various percentages related to the classification and gathering of webpages from different sources. The table shows that a total of 1,254,925 webpages were crawled, and that 15,361 relevant webpages and 2,122 relevant tweets were found. 

The system also demonstrated a high harvest rate of 81.70\%, indicating that a large number of pages were crawled relative to the number of relevant pages found. The table also shows that the system was able to classify a significant proportion of webpages as either coming from the clear web or the dark web, with the majority coming from the clear web. This was also reflected in the proportion of relevant webpages gathered from each source, with the majority coming from the clear web. In terms of the performance of individual spiders, the Ache and Sitemap spiders were responsible for the largest proportions of webpages gathered and relevant webpages found, respectively.

\begin{figure*}[hbt]
\centering

\begin{subfigure}[t]{0.75\textwidth}
    \includegraphics[width=\textwidth]{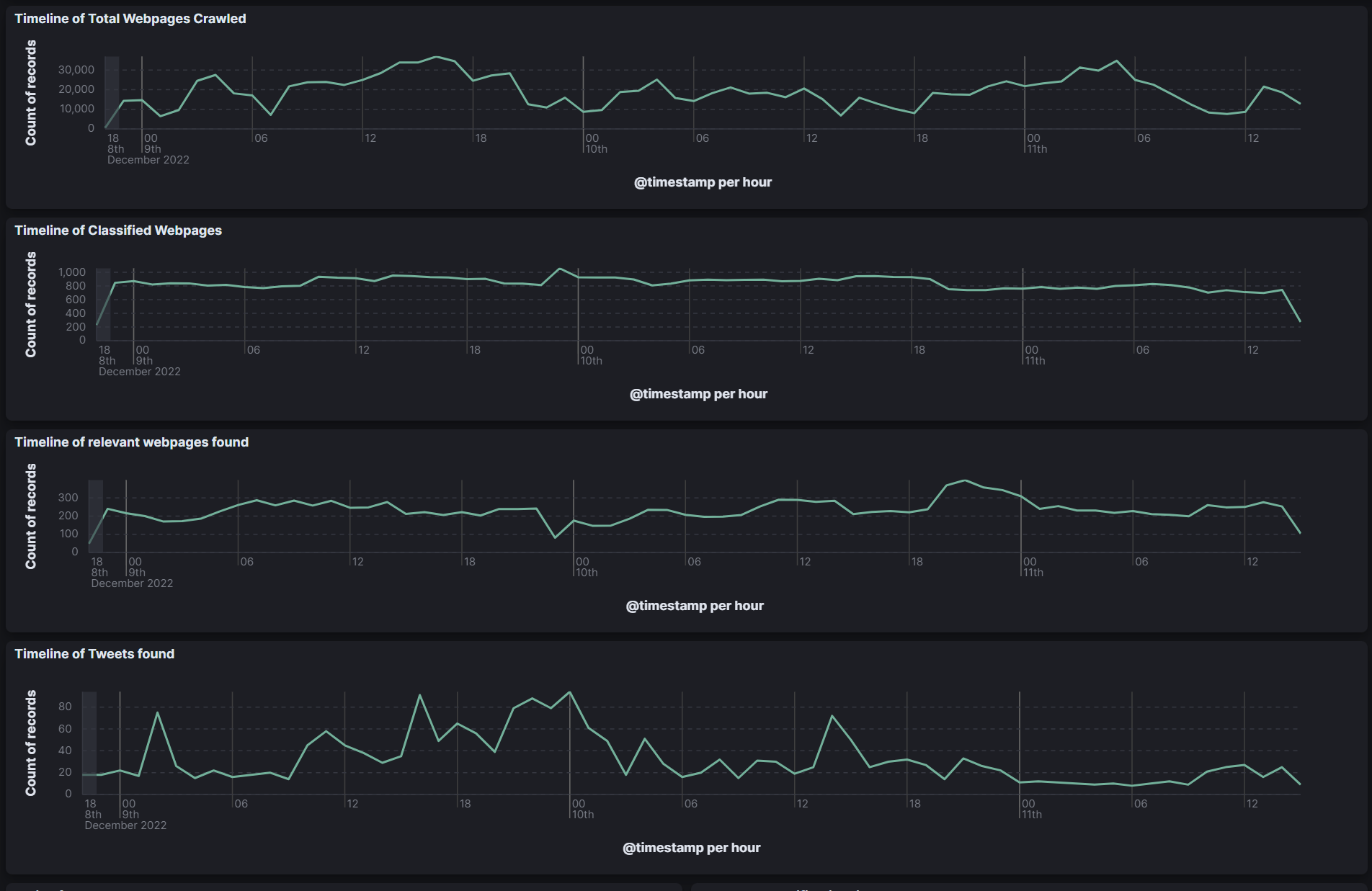}
    \caption{A Kibana dashboard displaying data on the activities of web crawlers over a period of three days.}
    \label{fig:Kibana_1}
\end{subfigure}

\vspace{0.3mm} 

\begin{subfigure}[t]{0.85\textwidth}
    \includegraphics[width=\textwidth]{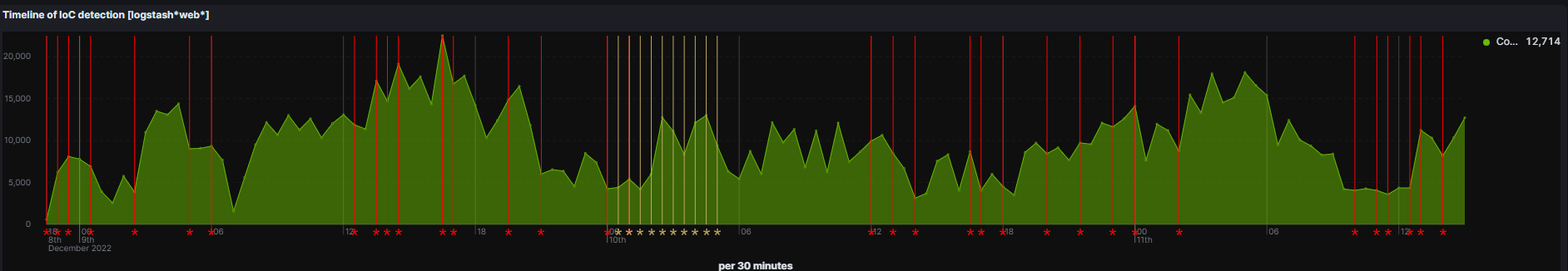}
    \caption{Timeline for IOC detection from Webpages}
    \label{fig:Kibana_2}
\end{subfigure}

\caption{Combined view of Kibana dashboard and Timeline for IOC detection.}
\label{fig:combined_Kibana}

\end{figure*}

Furthermore, Figure \ref{fig:Kibana_1} illustrates a dashboard displaying data for each hour of crawling for the clear web, dark web, and Twitter crawlers. The graph showcases the number of webpages and tweets crawled, as well as the number of relevant pages and tweets found. Figure \ref{fig:Kibana_2} presents a graph of IOCs (indicators of compromise) captured, using different colors of asterisk signs to indicate the resource used to capture each IOC. Overall, these results demonstrate the system's effectiveness in collecting and analyzing web data, and highlight the importance of using multiple spiders to maximize the number of relevant pages found.

\begin{table*}[hbt!]
\caption{System Metrics for Crawlers}
\label{tab:crawl_eval}    
\begin{tabular}{p{3.6in} p{2.69in}}
\hline\noalign{\smallskip}
Artifacts collected & Quantitative/Qualitative measurements\\
\noalign{\smallskip}\hline\noalign{\smallskip}
 Total crawled webpages & 1,254,925\\
 Relevant webpages found & 15,361 \\
 Relevant tweets found & 2,122 \\
 Percentage of webpages classified from each source & Clear Web -58.39 / Dark Web -41.61 \\
 Percentage of relevant webpages gathered from each source & Clear Web -73.71 / Dark Web -26.29 \\
 Percentage of webpages gathered by each spider & ache - 31.82/ sitemap -24.34/ ahmia -18.37/ wiki2 -24.33/ wiki1 -1.15\\
 Percentage of relevant webpages found  by each spider & ache - 29.67/ sitemap -44.04/ ahmia -12.92/ wiki2 -10.79/ wiki1 -2.58\\
 Harvest Rate (Total Pages Crawled/Number of Relevant Pages
) & 81.70 \\
\noalign{\smallskip}\hline
\end{tabular}
\end{table*}

\begin{table*}[hbt]
\caption{System Metrics for IOCs}
\label{tab:System_Metrics_for_IOCs}       
\begin{tabular}{p{3.8in} p{2.65in}}
\hline\noalign{\smallskip}
Artifacts collected & Quantitative measurements\\
\noalign{\smallskip}\hline\noalign{\smallskip}
 Total number of IOCs gathered & 163\\
 Number of unique IOCs gathered & 79\\
 Number of IOCs verified as valid & 69\\
 Time period used for verification of IOCs & 6 months\\
 Percentage of IOCs received from each resource & Twitter - 50.3/Clear Web - 29.4/Dark Web - 20.2\\
 Number of IOCs received from each resource & Twitter - 82/Clear Web - 48/Dark Web - 33\\
\noalign{\smallskip}\hline
\end{tabular}
\end{table*}

\begin{table*}[ht]
\caption{Sample of verified IOCs collected}
\label{tab:verified_IOCs_collected}       
\centering
\begin{tabular}{p{0.6in} p{0.3in} p{3.8in}  p{0.6in} p{0.6in}}
\hline\noalign{\smallskip}
Source  & Type & Sample IOCs collected & VirusTotal & Alienvault \\
\noalign{\smallskip}\hline\noalign{\smallskip}
Twitter & url & http://88[.]119[.]169[.]53/, & \quad \quad $\checkmark$ & \quad \quad $\times$\\
    Twitter    &  url  &   http://88[.]119[.]169[.]56/, & \quad \quad $\checkmark$ & \quad \quad $\times$\\
  Twitter      &   url   & http://168.100.8.160/, & \quad \quad $\checkmark$ & \quad \quad $\times$\\ 
   Twitter     &   url   &https://t.co/yYu1KoZvO1, & \quad \quad $\times$  & \quad \quad $\times$\\
    Twitter    &   url   & http://193.38.55.43/, & \quad \quad $\checkmark$ & \quad \quad $\checkmark$\\
    Twitter    &   url   & http://157.90.132.182/, & \quad \quad $\checkmark$ & \quad \quad $\checkmark$\\ 
    Twitter    &   url   & https://t.co/ynfw0e3dgC,  & \quad \quad $\times$  & \quad \quad $\times$\\
   Twitter     &   url   & https://nftuart.com/InvoiceTemplate.dotm & \quad \quad $\checkmark$ & \quad \quad $\times$\\ 
    Twitter    &  DNS  & wordpress-123380-0.cloudclusters.net & \quad \quad $\checkmark$ & \quad \quad $\checkmark$\\
   Twitter     & Hash & cd09bf437f46210521ad5c21891414f236e29aa6869906820c7c9dc2b565d8be & \quad \quad $\checkmark$ & \quad \quad $\checkmark$\\
Dark web & Hash & C2b8c65B0fBC9723E7af0EC5DD30746e77Ab3b65, & \quad \quad $\times$  & \quad \quad $\times$\\
    Dark web     & Hash & ABCD1234CDEF5678ABCD1234CDEF5678ABCD1234,  & \quad \quad $\times$  & \quad \quad $\times$\\
    Dark web     & Hash & Fbd055EEeA3b5a3459FeC6A8FAe631305b1079A0 & \quad \quad $\times$  & \quad \quad $\times$\\
    Dark web     & Hash & d282e137db2d55ae8fd3a299136f277e & \quad \quad $\checkmark$ & \quad \quad $\times$\\
    Dark web     & Hash & 192a8bf8a804e09670156b4bbb745387 & \quad \quad $\checkmark$ & \quad \quad $\times$\\
    Dark web     & Hash & 485b6e2bef303251789827d7829e3a3e & \quad \quad $\checkmark$ & \quad \quad $\checkmark$\\
Clear web & Hash & 50dbafed23e6e75d3f6313bf5480810a, & \quad \quad $\times$  & \quad \quad $\times$\\
    Clear web      & Hash & a95dad9c3d1b4b2b4ad6fd05961a1a3957500b2d  & \quad \quad $\times$  & \quad \quad $\times$\\
    Clear web      & Hash & 8fcaf491499ed2505d5a6523f561bbcc & \quad \quad $\checkmark$ & \quad \quad $\checkmark$\\
    Clear web      & Hash & 7593ec1357315431b04a17a55f01bd1295ca4b00ce8b910f8854a7e414e8f2cc & \quad \quad $\checkmark$ & \quad \quad $\checkmark$\\
     Clear web     & DNS & expiredaccessreviewnow.com & \quad \quad $\checkmark$ & \quad \quad $\checkmark$\\
     Clear web     & DNS &  bafybeicrq42t3uoi53hf2hhntwq74hfapj5rutrp6ejlidohaghypibnyy.ipfs.dweb.link & \quad \quad $\checkmark$ & \quad \quad $\checkmark$\\
\noalign{\smallskip}\hline
\end{tabular}
\end{table*}

\subsubsection{IOC analysis}
Table \ref{tab:System_Metrics_for_IOCs} presents the results of the system's performance in collecting IOCs from open sources. Over the course of three days, a total of 163 IOCs were gathered, with 79 of them being unique. We verified the existence of identified IOCs on two different platforms (Alienvault and VirusTotal) and rechecked their status, especially those initially classified as Non-IOCs in the previous six-month verification cycle.
Based on the results obtained from this verification cycle, we found that 69 out of the 79 initially obtained IOCs are valid. If a potential threat (IOC) cannot be found in the databases, two possible scenarios arise. Firstly, the entity in question may not constitute a threat, indicating no involvement in any attacks, and consequently, it may not appear in the databases. Alternatively, it could represent a novel threat that has been identified by the TSTEM system but has not yet been integrated into the existing databases. This aligns with the primary objective behind the development of TSTEM – to identify and respond to emerging threats that have not been documented in current databases. The "Time period used for verification of IOCs" indicates the verification period for IOCs. The table also illustrates the percentage and number of IOCs received from each resource, with Twitter being the most prominent source, followed by the clear web and then the dark web. Specifically, the system was able to capture a relatively large proportion of IOCs from Twitter, with approximately 50\% of the total number of IOCs being collected from this source.

Furthermore, the system captured a significant proportion of IOCs from the clear web (30\%), and a smaller proportion from the dark web (20\%). Table \ref{tab:verified_IOCs_collected}  presents a sample of the Indicators of Compromise that were gathered and validated during the study. The authenticity of these IOCs was confirmed on two platforms: VirusTotal and Alienvault. The table consists of five columns: "Source," "Type," "Sample IOCs Collected," "VirusTotal," and "Alienvault." The "Source" column indicates the origin of the IOCs, which may be Twitter, the dark web, or the clear web. The "Type" column specifies the nature of the IOC, whether it is a URL, a hash, or another type of indicator. The "Sample IOCs Collected" column provides a selection of IOCs gathered from each source and of each type. The "VirusTotal" and "Alienvault" columns denote the platforms used to verify the existence of the IOCs. The results obtained during verification are marked with a tick if found or a cross if not found on the respective platform.

These findings suggest that the system was effective at collecting a significant number of IOCs from various sources, with a relatively high proportion of unique indicators.  The higher proportion of IOCs received from Twitter may be due to the large volume of data available on this platform, as well as the potential for real-time information gathering. However, it is possible that the poor performance of the web crawlers in finding IOCs (indicators of compromise) is due to a combination of factors, including the design of the crawlers and the overall complexity of the task. To enhance the performance of the crawler in identifying IOCs, a potential solution may involve reconfiguring the crawler to more effectively target and gather pertinent information specific to the task under consideration.

\section{Discussion}
The majority of cyber threat intelligence (CTI) is currently stored in data warehouses or data lakes, creating technical bottlenecks and opaque CTI feeds. To bridge this gap, our research introduces TSTEM, a platform that incorporates several innovative approaches to optimize the efficacy and efficiency of CTI collection in the wild. TSTEM incorporates AI-based focus crawlers that use custom domain language models to enable targeted CTI collection, indexing, and IOC extraction from clear, deep, and social web sources. Moreover, TSTEM uses a containerized architecture that is capable of managing computationally demanding data pipelines, ensuring efficient and scalable CTI collection. Finally, we demonstrate the advantages of deploying CTI tools as infrastructure as code, highlighting the potential for quick and consistent deployment. In this section, we present a comparative analysis of our research outcomes and performance metrics relative to those of other contemporary state-of-the-art models operating within the same domain.

Numerous studies have focused on the extraction of appropriate Indicators of Compromise from OSINT sources, incorporating Artificial Intelligence algorithms into their methodologies. However, the majority of these studies predominantly rely on statistical approaches or elementary Machine Learning and Deep Learning algorithms. In contrast, our proposed framework employs a multi-level extraction process, which combines multiple AI-Natural Language Processing (NLP) algorithms tailored to the data inputs gathered from web crawlers.

Our research results have demonstrated notable success in comparison to existing models, such as the one proposed by Pham et al. \cite{pham2018phishing}. They employed a neuro-fuzzy model, referred to as Fi-NFN, which achieved an accuracy of approximately 98.36\%. While the accuracy of their model is indeed comparable to the results obtained by our framework, there are several aspects in which our approach outperforms the Fi-NFN model. Some of the facts include that the paper only addresses the detection of phishing websites, not the extraction of IOCs from the content of the websites. Additionally, it does not provide a re-deployable platform or framework powered by MLOps. We are using SOTA algorithms as part of our framework and advanced technical components like BentoML as part of the framework.

Our findings reveal a distinct improvement in precision when measured against the research carried out by Ghazi et al. \cite{ghazi2018supervised}. Their work achieved a precision of 70\% by employing a combination of a Named Entity Recognition algorithm (Continuous Random Field) and a rule-based algorithm. In contrast, our study utilizes a more advanced NER algorithm, leading to enhanced precision of 83\%. However, further optimization of the algorithm may be required to maintain the upward trajectory in precision. Similarly, Mittal et al. \cite{mittal2019cyber} utilized a Named Entity Recognition model in conjunction with relationship extraction and knowledge graph representation, achieving an accuracy of 81.5\%. Despite the integration of multiple natural language processing techniques in their methodology, the resulting accuracy was lower than that attained in our study, which is around 98\%. To improve accuracy in future research, further optimization of the NER model and associated techniques may be necessary.

In \cite{al2022cyber}, authors applied PCA for dimension reductionality, used a basic DNN model, and obtained 98\% accuracy only. An RNN-based classifier for classifying web page contents was used in \cite{kannegantirecurrent} and reached accuracy of 85\% , whereas our model achieved 95\%. In \cite{tanvirul2022cyner}, despite the use of heuristic methods, a transformer-based model (XLM-Roberta), and Spacy algorithms, the combined precision, recall, and F1 score were only 75.30\%, 78.07\%, and 76.66\%, respectively. In contrast, our study achieved a precision of 83\%, recall of 85\%, and an F1 score of 84\%.Vlachos et al. in \cite{vlachos2022saint} employed statistical methods with a rule-based algorithm, achieving good performance in extracting IOCs. The study in \cite{nunes2016darknet} experimented with Naive Bayes, Random Forest, Support Vector Machine, and Logistic Regression, reaching a maximum recall of 92\% and precision of 82\%, while our model achieved a precision of 83\%, recall of 85\%, and an F1 score of 84\%.

Some of the significant impact of our approach and methodologies includes:

\begin{enumerate}
\item Enhanced threat detection and quicker response times achieved through continuous monitoring and data extraction.
\item Real-time collaboration and knowledge-sharing capabilities for improved decision-making.
\item Comprehensive Threat Intelligence Analysis, involving continuous monitoring of multiple external threat feeds, as well as data aggregation and analysis from diverse sources.
\item The establishment of an easily deployable and maintainable platform, all made possible without manual intervention,  through the use of Infrastructure as Code (IaC) methods and MLOps for managing the Cyber Threat Intelligence infrastructure.
\item These outcomes underscore the robust architectural design of our framework, its utilization of advanced natural language processing algorithms delivering superior accuracy and performance, and its straightforward and uninterrupted deployment procedures.
\item We employed BentoML to offer Model as a Service, addressing the challenge of employing machine \item learning models as a service in an innovative manner, a path less explored in prior studies.
The adoption of Infrastructure as Code and MLOps approaches can significantly boost the efficiency and effectiveness of CTI infrastructure deployment and maintenance.

\end{enumerate}

In detail, compared with other studies, our study used a continuous multi-level extraction method with a transformer-based algorithm for classification, a transformer-based entity extraction algorithm, and a heuristics-based logic system. The study attained optimal performance for three distinct models, namely BERT (short-text classifier) with 98\% accuracy, Longformer (web page classifier) with 95\% accuracy, and BERT (Entity recognition model) with 98\% accuracy. The above results indicate that our framework has a more robust architecture, utilizes advanced natural language processing algorithms with higher accuracy and performance, and has a straightforward continuous deployment process. 

In Table \ref{tab:Comparison of results CTI platforms with AI}, the efficacy of diverse AI platforms and approaches within the Cyber Threat Intelligence domain is systematically compared, emphasizing key performance indicators such as accuracy, precision, recall, and F1-score. The data, derived from several research papers, elucidates the varied metrics achieved by these platforms. Notably, the table highlights the superior performance of the TSTEM platform across several classifiers and IOC extractors, underscoring its proficiency in CTI operations.

Furthermore, our proposed framework employs MLOps and Infrastructure as Code (IaC) methodologies to facilitate the deployment and upkeep of the Cyber Threat Intelligence infrastructure. The IaC methodology allowed for the creation and management of the entire infrastructure as a code base, thereby simplifying the processes of deployment and maintenance for the system. To demonstrate the framework's effectiveness, we used Ansible and Terraform as part of our cloud computing paradigm to deploy CTI tools. Ansible is an open-source tool that helps with IT configuration management, while Terraform is a tool that is used for efficiently and safely building, changing, and versioning infrastructure. By using these tools, we were able to show how infrastructure as code can be used to easily deploy and manage CTI tools without the need for manual intervention. Overall, using the Infrastructure as Code and MLOps approach can make the deployment and maintenance of CTI infrastructure more efficient and effective. Furthermore, we used BentoML to provide a Model as a Service as part of the framework, which addresses the challenge of using machine learning models as a service that has yet to be fully explored in earlier studies.



\begin{table*}[h!]
\centering 
\caption{Comparison of results from core papers with AI.}
\label{tab:Comparison of results CTI platforms with AI} 

\centering
\begin{tabular}{p{1.5in} p{1.5in} p{0.9in}  p{0.9in} p{0.9in}}
\hline\noalign{\smallskip}
Citation  & Accuracy & Precision & Recall & F1-score \\
\noalign{\smallskip}\hline\noalign{\smallskip} 
\noalign{\smallskip}\hline\noalign{\smallskip}
\cite{dutta2020overview} & 96.6\% &97\% & 97\%& 97\%\\
\cite{ghazi2018supervised} & &69\% & 56\% & 62\%\\
\cite{mittal2019cyber} & 81.5\% & 84\% & 24\% & \\
\cite{al2022cyber} &98\% & & & \\
\cite{kannegantirecurrent} & 85\%&  & & \\
\cite{tanvirul2022cyner} & & 75.30\% &78.07\% & 76.66\% \\
\cite{nunes2016darknet} & & 82\%&92\% &  \\
\cite{koloveas2021intime} & 95\% & 61\% & 73\% & 64\%\\
\cite{zhao2020timiner}  & CTI: 84\% IOC: 94\% & & & \\
\cite{guarascio2022boosting} &  & & & 97.1\%\\
\cite{10.1007/978-3-031-36096-1_4} & 80\% \\
\textbf{TSTEM} (our platform) &  & & & \\
- Sentence classifier& 98\%& 98\%& 96\%&  97\%\\
- CTI web page classifier&95\% &  96\%&  92\%&  94\%\\
- IOC extraction (NER)& 98\%&  83\%&   85\%&  84\% \\
\noalign{\smallskip}\hline
\end{tabular}
\end{table*}

\section{Conclusion and future work}

Automatically extracting Cyber Threat Intelligence (CTI) from open sources is a rapidly expanding defence concept that provides enhanced resilience to cyber attacks. However, identifying and extracting valuable Indicators of Compromise (IOCs) is a complex challenge that demands the seamless integration of multiple technology stacks, tools, and pipelines. Consequently, obtaining valuable threat intelligence from open sources in the wild remains both a bottleneck and a black box. To bridge this gap, we propose a platform (TSTEM) that combines artificial intelligence (AI), web crawling, MLOps, and cloud computing to target CTI identification and extraction tasks. TSTEM is a robust and efficient system that incorporates focused crawling along with MLOps and Infrastructure as Code (IaC) components.

Twitter's free API allows researchers and developers to access a small, random selection of tweets from the entire platform, representing approximately 1\% of all tweets. When conducting hashtag-based searches using this API, the 1\% limitation still applies, but it is not specifically defined for tweets within a particular hashtag. Instead, it broadly pertains to the aggregate volume of tweets spanning all hashtags and subjects available through the free API. Performing hashtag-based searches using Twitter's free API means that this limitation affects our method for collecting data. 

There are several enhancements that we plan to implement in future versions of the framework to improve its performance and effectiveness. These include:

\begin{enumerate}
\item Improving the focused crawling logic to more effectively locate indicators of compromise (IOCs). This may involve analyzing the design and capabilities of the web crawlers and identifying areas for improvement.
\item Enhancing the accuracy of the models through the use of more refined training data.
\item Automating the generation of STIX objects from the IOCs and context captured from the CTI data.
\item Introducing pipeline automation, continuous integration, continuous delivery, continuous testing, and parallel processing with MLOps to enable users to train models on their own data and facilitate continuous re-training and deployment of the model.
\item Focusing on intelligence beyond atomic IOCs, which are relatively easy for threat actors to overcome (low pain), to better address the overall threat landscape.
\end{enumerate}

Overall, our framework is designed to be a robust and effective tool for collecting and analyzing web data, with high-end natural language processing algorithms and a straightforward continuous deployment process. By implementing the aforementioned enhancements, we aim to further improve its performance and usefulness for cybersecurity professionals.\\

\textbf{Declaration of Competing Interest}\\
The authors declare that there are no known conflicting financial interests or personal relationships that may seem to have impacted the findings presented in this paper.
\paragraph{}

\textbf{CRediT authorship contribution statement}\\
\textbf{Prasasthy Balasubramanian:} Writing - original draft; Writing - review and editing, Conceptualization, Data curation; Formal analysis; Software; Investigation; Methodology, Validation. \textbf{Sadaf Nazari:} Writing - original draft, Data curation; Formal analysis; Software; Validation. \textbf{Danial Khosh Kholgh}: Writing - original draft, Data curation; Formal analysis; Software; Visualization, Validation. \textbf{Alireza Mahmoodi: } Data curation, Methodology. \textbf{Justin Seby: } Data curation, Methodology. \textbf{Panos Kostakos: }Writing - original draft; Writing - review and editing, conceptualization, funding acquisition, Investigation; Methodology.

\paragraph{}
\section*{Acknowledgments}
This work was funded by the European Commission grants IDUNN (grant no. 101021911) and Academy of Finland 6Genesis Flagship (grant no. 318927).
\paragraph{} 

\bibliographystyle{cas-model2-names}

\bibliography{cas-refs}


\bio{}
\textbf{Prasasthy Balasubramanian} is a Ph.D. researcher at the Center for Ubiquitous Computing, University of Oulu, Finland. She received the MS degree of Science in Data Science from Liverpool John Moores University, UK. Her current research interests include cyber threat intelligence, deep learning, Conversational AI, Explainable AI and Generative AI.
\endbio
\bio{}
\textbf{Sadaf Nazari} is currently pursuing a Master's degree in Computer Science and Engineering at the University of Oulu, Finland  
\bio{}
\textbf{Danial Khosh Kholgh} is currently pursuing a Master's degree in Computer Science and Engineering at the University of Oulu, Finland.
\endbio
\bio{}
\textbf{Alireza Mahmoodi} is currently pursuing a Master's degree in Computer Science and Engineering at the University of Oulu, Finland.
\endbio
\bio{}
\textbf{Justin Seby} is currently pursuing a Master's degree in Computer Science and Engineering at the University of Oulu, Finland .
\endbio
\bio{}
\textbf{Panos Kostakos} serves as a Senior Research Fellow at the Center for Ubiquitous Computing in the University of Oulu, Finland. His interdisciplinary research concentrates on the synergistic fusion of artificial intelligence, information security, and security orchestration with a primary focus on developing adaptive, autonomous, and cognitive cybersecurity defence mechanisms. He leads the research group Cyber Security Informatics (CSI).
\endbio

\end{document}